\documentclass[showpacs,aps,floatfix,pra,reprint,superscriptaddress]{revtex4-1}
\usepackage{footnote}
\usepackage{enumitem}
\makesavenoteenv{tabular}

\usepackage{scrextend}
\usepackage{xfrac}
\usepackage{amssymb}
\usepackage{braket}
\usepackage{graphicx}
\usepackage{verbatim}
\usepackage{mhchem}
\usepackage{etoolbox}
\usepackage{amsfonts} 
\usepackage{amsmath}  
\usepackage[utf8]{inputenc} 
\usepackage{mathtools}
\usepackage{soul,xcolor,colortbl}
\usepackage{hyperref} \hypersetup{colorlinks=true,citecolor=blue,linkcolor=blue,urlcolor=blue} 
\usepackage{tabularx}
\usepackage{bm}    
\usepackage{array}
\usepackage{color} 
\usepackage{flexisym}
\usepackage[super]{nth}
\newcolumntype{L}[1]{>{\raggedright\let\newline\\\arraybackslash\hspace{0pt}}m{#1}}
\newcolumntype{C}[1]{>{\centering\let\newline\\\arraybackslash\hspace{0pt}}m{#1}}
\newcolumntype{R}[1]{>{\raggedleft\let\newline\\\arraybackslash\hspace{0pt}}m{#1}}
\usepackage{comment}

\usepackage[autostyle]{csquotes}
\usepackage{rotating}

\usepackage{scalerel}[2016/12/29]
\usepackage{multirow}
\usepackage{siunitx}
\usepackage{adjustbox}
\usepackage{blindtext}
\usepackage{adjustbox}

\newcommand{\SC}[1]{{ #1}}
\SetLabelAlign{Center}{\hfil#1\hfil}
\SetLabelAlign{CenterWithParen}{\hfil(\makebox[1.0em]{#1})\hfil}

\def\sP{{\substack{\scalebox{0.6}{P}}}}
\def\sV{{\substack{\scalebox{0.6}{V}}}}
\def\sB{{\substack{\scalebox{0.6}{B}}}}

\def\sEQ{{\substack{\scalebox{0.6}{eq}}}}
\def\sEE{{\substack{\scalebox{0.6}{e}}}}

\def\sphelec{{\substack{\scalebox{0.6}{ph+elec}}}}  

\def\selectr{{\substack{\scalebox{0.6}{elec}}}}  

\def\svib{{\substack{\scalebox{0.6}{vib}}}}   
\def\szero{{\substack{\scalebox{0.6}{0}}}}    

\def\AFLOW{{\small AFLOW}}

\def\VASP{{\small VASP}}
\def\SPBE{{\small PBE}}
\def\SPBESOL{{\small PBEsol}}
\def\SXC{{\small XC}}
\def\QHA{{\small QHA}}

\def\QHAPPP{{\small QHA3P}}
\def\SCQHA{{\small SC-QHA}}

\def\SRMSRD{{\small RMSrD}}
\def\EOS{{\small EOS}}

\def\EXP{{\mathrm{exp}}}
\def\Gruneisen{{Gr\"{u}neisen}}

\definecolor{pranab_green}{rgb}{0.31,0.53,0.10}
\definecolor{pranab_red}{rgb}{0.85,0.23,0.11}

\def\DUKEMEMS{Department of Mechanical Engineering and Materials Science, Duke University, Durham, North Carolina 27708, USA}
\def\DUKECMS{Center for Materials Genomics, Duke University, Durham, NC 27708, USA}
\def\UNTPHYS{Department of Physics and Department of Chemistry, University of North Texas, Denton TX, USA}
\def\CMUPHYS{Department of Physics, Central Michigan University, Mount Pleasant, MI 48858, USA}
\def\FHI{Fritz-Haber-Institut der Max-Planck-Gesellschaft, 14195 Berlin-Dahlem, Germany}

\makeatletter \renewcommand\frontmatter@abstractwidth{\dimexpr\textwidth\relax} \makeatother 

\begin{document}

\title{AFLOW-QHA3P: Robust and automated method \\ to compute thermodynamic properties of solids}

\author{Pinku Nath}\affiliation{\DUKEMEMS}\affiliation{\DUKECMS}
\author{Demet Usanmaz}\affiliation{\DUKEMEMS}\affiliation{\DUKECMS}
\author{David Hicks}\affiliation{\DUKEMEMS}\affiliation{\DUKECMS}
\author{Corey Oses}\affiliation{\DUKEMEMS}\affiliation{\DUKECMS}
\author{Marco Fornari}\affiliation{\CMUPHYS}\affiliation{\DUKECMS}
\author{Marco Buongiorno Nardelli}\affiliation{\UNTPHYS}\affiliation{\DUKECMS}
\author{Cormac Toher}\affiliation{\DUKEMEMS}\affiliation{\DUKECMS}
\author{Stefano Curtarolo}\email[]{stefano@duke.edu}\affiliation{\DUKEMEMS}\affiliation{\DUKECMS}\affiliation{\FHI}

\date{\today}


\begin{abstract}
\noindent
Accelerating the calculations of finite-temperature thermodynamic
properties is a major challenge for rational materials design.
Reliable methods can be quite expensive, limiting their effective
applicability in autonomous high-throughput workflows.
Here, the 3-phonons quasi-harmonic approximation ({\small QHA}) method is introduced, 
requiring only three phonon calculations to obtain a thorough characterization of the material.
Leveraging a Taylor expansion of the phonon frequencies around the equilibrium volume, the method efficiently resolves 
the volumetric thermal expansion coefficient, specific heat at constant pressure, the enthalpy, and bulk modulus.
Results from the standard {\small QHA} and experiments corroborate the procedure,
and additional comparisons are made with the recently developed
self-consistent {\small QHA}.
The three approaches --- 3-phonons, standard, and self-consistent
{\small QHA}s --- are all included within the automated, open-source
framework {\small AFLOW}, 
allowing automated determination of properties with various implementations within the same framework. 

\end{abstract}

\maketitle
\section*{\label{sec:Intro}Introduction}

\begin{figure*}[]
\centering
\includegraphics*[width=0.9\textwidth]{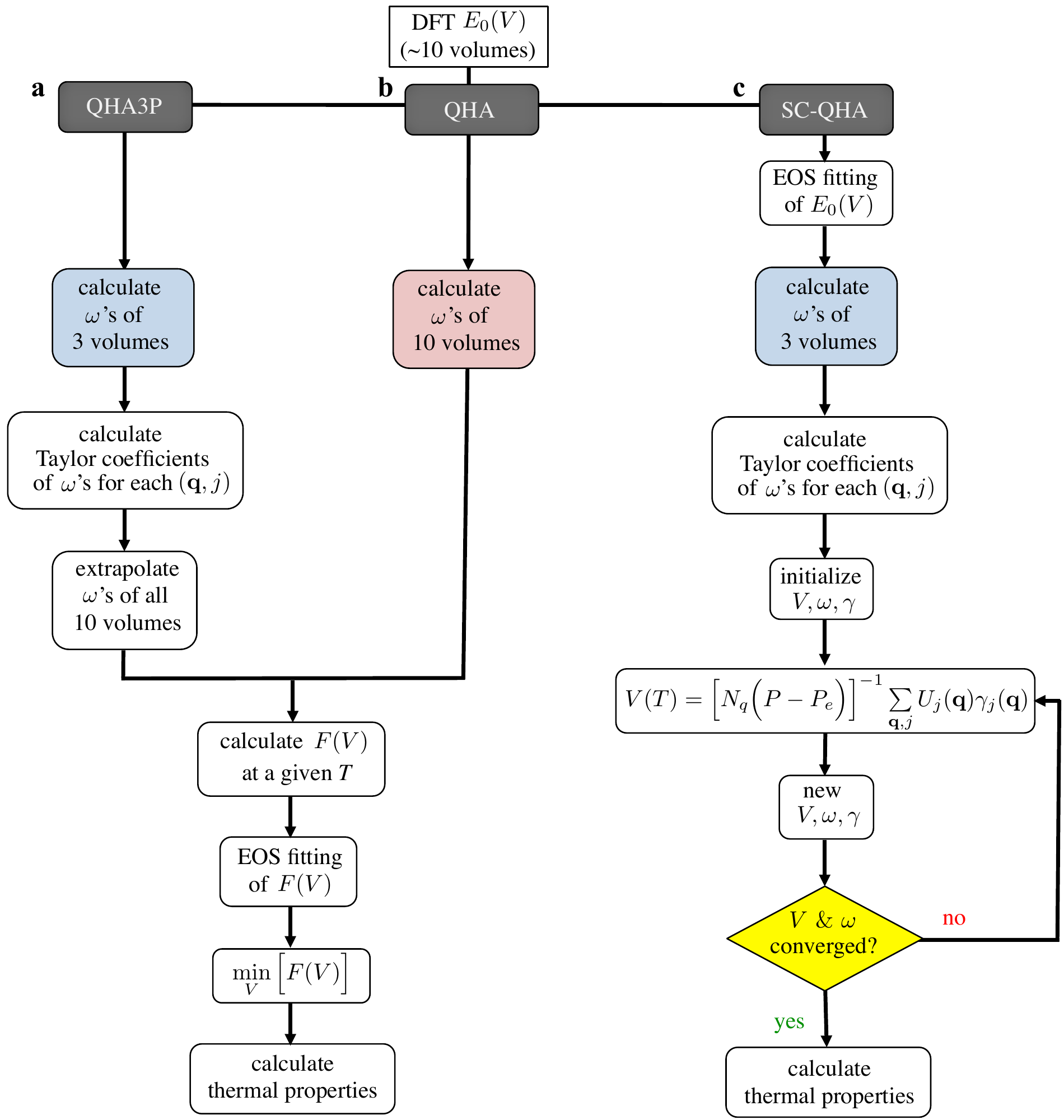}
\vspace{-2mm}
\caption{\small The workflows for ({\textbf a}) \QHAPPP\ compared with ({\textbf b}) \QHA\ and ({\textbf c}) \SCQHA.
The most time-consuming step of \QHA\ is highlighted in red, while the time-saving steps of the other two methods are highlighted in blue.}
\label{Fig:fig1}
\end{figure*}

\par
Reliable and efficient computational methods are needed to guide time-consuming and laborious experimental searches, thus accelerating materials design.
Implementing effective methods within automated frameworks such as \AFLOW~\cite{curtarolo:art65,curtarolo:art75,curtarolo:art92,curtarolo:art127,curtarolo:art128} facilitates the calculation of thermodynamic properties for large materials databases.
There are several computational techniques to characterize the temperature dependent properties of materials, each with varying accuracy and computational cost.
Techniques such as {\it ab initio}  molecular dynamics~\cite{Yu_SR_2016,Kresse_PRB_1993,Sarnthein_PRB_1996,Zhang_APL_2015,Jesson_APL_2015} and the stochastic self-consistent harmonic approximation~\cite{Errea_PRL_2013,Errea_PRB_2014} give accurate results for the temperature dependent properties of materials.
Although these methods are highly accurate, the treatment of anharmonicity requires the consideration of many large distorted structural configurations, making them computationally prohibitive for screening large materials sets.
Other methods including the Debye-\Gruneisen\ model~\cite{BlancoGIBBS2004,Poirier_Earth_Interior_2000} or Machine Learning approaches~\cite{curtarolo:art124,curtarolo:art136} require less computational resources, but often struggle to predict properties such as the \Gruneisen\ parameter with reasonable accuracy~\cite{curtarolo:art96,curtarolo:art115}. 
The \underline{q}uasi-\underline{h}armonic \underline{a}pproximation (\QHA)~\cite{Liu_Cambridge_2016,Wang_ACTAMAT_2004,Duong_jap_2011} balances accuracy and computational cost for calculating temperature and pressure dependent properties of materials.

\par
In its standard formulation, \QHA\ also remains too expensive for automated screening \cite{curtarolo:art114}.
\QHA\ requires many independent phonon spectra calculations, obtained by diagonalizing the dynamical matrices giving eigenvectors (modes) and eigenvalues (energies)~\cite{Ziman_Oxford_1960,ThermoCrys,Dove_LatDynam_1993,PhysPhon}.
The dynamical matrix can be constructed either with \enquote{linear response}~\cite{PhysPhon,BaroniRMP2001} or the \enquote{finite displacements} method~\cite{PhysPhon,Axel_RMP,curtarolo:art125}.
Despite its low computational demands, linear response does not perform well at high temperatures where anharmonicity can be large.
The finite displacements method can be easily integrated with a routine that computes forces, making the preferred method for high-throughput calculations.
However, it is still computationally expensive for
(i) low symmetry crystals, 
(ii) materials with large atomic variations leading to complicated optical branches, and
(iii) metallic systems having long-range force interactions requiring large supercells.
In order to make \QHA\ more suitable for automated screening, it is necessary to reduce the number of required phonon spectra.

\par
Recently, the \underline{s}elf-\underline{c}onsistent \underline{q}uasi-\underline{h}armonic \underline{a}pproximation~\cite{Huang_cms_2016} (\SCQHA) has been developed.
It self-consistently minimizes the external and internal pressures.
The method requires spectra at only two or three volumes, while the frequency-volume relationship is determined using a Taylor expansion.
It is computationally efficient and almost five times faster than \QHA.
Results agree well with experiments at low temperatures, although some deviations are observed at high-temperature for the tested systems~\cite{Huang_cms_2016}.

\par
In this article, the \underline{q}uasi-\underline{h}armonic \underline{a}pproximation \underline{3}-\underline{p}honons (\QHAPPP) method is introduced.
It calculates the phonon frequencies around equilibrium for only three different volumes, and performs a Taylor expansion to extrapolate the phonon frequencies at other volumes.
The \QHAPPP\ approach drastically reduces the computational cost and achieves consistency with experiments, allowing automated materials' property screening without compromising accuracy.
Similar to \QHA, \QHAPPP\ minimizes the Helmholtz free energy with respect to volume for each temperature.
The calculation of the thermodynamic properties, and the temperature dependent electronic contribution to the free energy, are the same as in \QHA, enabling unrestricted screening of all types of materials.
The \QHA, \SCQHA, and \QHAPPP\ methods are all implemented within \AFLOW~\cite{curtarolo:art114,curtarolo:art65,curtarolo:art75,curtarolo:art92,curtarolo:art127,curtarolo:art128}.
The performance of \QHAPPP\ with two different e\underline{x}change-\underline{c}orrelation (\SXC) functionals is investigated.


\section*{\label{sec:Comp-detalis}Computational details}

\par
The thermal properties of materials at finite temperatures are calculated from the Helmholtz free energy $\left( F \right)$, which depends on temperature $\left( T \right)$ and volume $\left( V \right)$.
Neglecting the electron-phonon coupling and magnetic contributions, $F$ can be written as the sum of three additive contributions~\cite{Liu_Cambridge_2016,Wang_ACTAMAT_2004,Duong_jap_2011}:

\begin{equation}\label{Eq:Ftot}
F\left(V,T\right)=E_\szero\left(V\right)+F_\svib\left(V,T\right)+F_\selectr\left(V,T\right),
\end{equation}

\noindent
where $E_\szero$ is the total energy of the system at $0$~K without any atomic vibrations, $F_\svib$ is the vibrational free energy of the lattice ions,
and $F_\selectr$ is the finite temperature electronic free energy due to thermal electronic excitations.


\vspace{5mm}
\noindent\textbf{\QHA\ Methodology.}
%
\QHA\ enables the calculation of $F_\svib$ via the harmonic approximation and includes anharmonic effects in the form of volume dependent phonon frequencies.
$F_\svib$ is given by~\cite{curtarolo:art114,Duong_jap_2011,Wang_ACTAMAT_2004,Liu_Cambridge_2016}:

\begin{widetext}
\begin{equation}\label{Eq:Fvib}
F_\svib\left(V,T\right)=\frac{1}{N_q}\sum_{{\bf q}, j}\left( \frac{\hbar\omega_{j}({\bf q})}{2}+k_{\sB}T{\rm ln}\left[1-\EXP\left(-\frac{\hbar\omega_{j}(\bf q)}{k_{\sB}T}  \right)\right] \right), 
\end{equation}
\end{widetext}

\noindent
where $\hbar$ and $k_\sB$ are the Planck and  Boltzmann constants, and $\omega_{j}(\bf q)$ is the volume dependent phonon frequency (at ${\bf q}, j$).
The (${\bf q}, j$) comprises both the wave vector {\bf q} and phonon branch index $j$.
$N_q$ is the total number of wave vectors.

\par
Although $F_\selectr \left(  V,T \right)$ is negligible for wide band gap materials, its contribution is required for metals and narrow band-gap systems.
$F_\selectr \left(  V,T \right)$ is calculated as~\cite{Wang_ACTAMAT_2004,Arroyave_ACTAMAT_2005,Liu_Cambridge_2016}:
\begin{equation}\label{Eq:Fele}
\begin{split}
F_\selectr \left(V,T\right)=& U_\selectr \left(V,T\right) - TS_\selectr \left(V,T\right), \\
U_\selectr \left(V,T\right)=&\int_{0}^{\infty}  n_\selectr \left(\epsilon \right)f\left(\epsilon\right)\epsilon {\rm d}\epsilon - \int_{0}^{E_{\rm F}} n_\selectr \left(\epsilon\right)\epsilon {\rm d}\epsilon, \\
S_\selectr \left(V,T\right)=&-k_{\sB}\int_{0}^{\infty} n_\selectr \left(\epsilon \right) [f\left(\epsilon\right){\rm ln}\left(f\left(\epsilon\right)\right)+\\
       &+ \left(1-f\left(\epsilon\right)\right){\rm ln}\left(1-f\left(\epsilon\right)\right)]{\rm d}\epsilon ,
\end{split}
\end{equation}

\noindent
where $U_\selectr \left(  V,T \right)$ and $S_\selectr \left(  V,T \right)$ are the temperature dependent parts of the electronic internal energy and the electronic entropy respectively, $n_\selectr(\epsilon)$ is the density of states at energy $\epsilon$, $f(\epsilon)$ is the Fermi-Dirac distribution, and $E_{\rm F}$ is the Fermi energy.


\par
The \QHA\ method requires at least ${\sim}10$ $E_\szero$ and ${\sim}10$ phonon calculations $\left( F_{\svib} \right)$ in order to obtain a good fit for the equation of state (\EOS).
\QHA\ is generally implemented using isotropic volume distortions, although its implementation in \AFLOW\ can also include anisotropic effects by considering $F$ as a function of direction-dependent strain, {\it e.g.} along principal directions~\cite{Tohei_JAP_2016,Hermet_JPCC_2014}.
The calculated $E_\szero$ and $F_{\svib}$ are fitted to an \EOS\ ({\it e.g}., Birch-Murnaghan \EOS~\cite{Fu_PRB_1983}).
The equilibrium volume $\left( V_\sEQ \right)$ at a given temperature is determined by minimizing $F$ with respect to $V$ at a given $T$, $\left(\partial F/\partial V\right)_{\scalebox{0.6}{T}}{=}0$ (Figure~\ref{Fig:fig1}{\bf b}).
A more detailed description of $F{-}V$ interpolation and the calculation of different energy terms is discussed in Ref.~\cite{curtarolo:art114}.


The thermodynamic properties --- constant volume specific heat $\left( C_{\sV} \right)$, constant pressure specific heat $\left( C_{\sP} \right)$, average \Gruneisen\ parameter $\left( {\bar\gamma} \right)$, and volumetric thermal expansion coefficient $\left( \alpha_\sV \right)$~\cite{curtarolo:art114,Duong_jap_2011,Wang_ACTAMAT_2004,Liu_Cambridge_2016} --- are calculated according to the following definitions:

\begin{eqnarray}
C_{\sV}  & = & k_{\sB}\sum_{{\bf q},j} \left(\frac{\hbar\omega_{j}({\bf q},V_\szero)}{k_{{\sB}}T}\right)^2 \frac{\EXP\left({\frac{\hbar\omega_{j}({\bf q},V_\szero)}{k_{{\sB}}T}}\right)}{\left(\EXP{\left(\frac{\hbar\omega_{j}({\bf q}, V_\szero)}{k_{{\sB}}T}\right)}-1\right)^2}
\label{Eq:Cp}\\[5pt]
C_{\sP}  & = & C_{\sV}+ V_{\sEQ}TB\alpha_\sV^2, \label{Eq:Cv} \\[5pt]
{\bar \gamma}  & = & \frac{\sum_{{\bf q},j} \gamma_{j}({\bf q}) c_{j}({\bf q})}{ \sum_{{\bf q},j} c_{j}({\bf q})}, \label{Eq:gammabar} \\[5pt]
\alpha_\sV  & = & \frac{C_{\sV}{\bar \gamma}}{BV_{\sEQ}}, \label{Eq:alphav}
\end{eqnarray}

\noindent
where $\omega_{j}({\bf q}, V_\szero)$ is the frequency of phonon mode (${\bf q},j$) at relaxed volume ($V_{\szero}$), and $c_{j}(\bf q)$ and $\gamma_{j}(\bf q)$ are the specific heat capacity at constant volume and mode \Gruneisen\ at (${\bf q},j$). 
The definitions of the bulk modulus ($B$) and mode \Gruneisen\ parameter, ($\gamma_{j}(\bf q)$), are

\begin{eqnarray}
B &  = &V_{{\sEQ}}\left(\frac{\partial^2 F}{\partial V^2}\right)_T, \label{Eq:B-qha} \\
\gamma_{j}(\bf q)  & = & -\frac{V_\szero}{\omega_{j}({\bf q}, V_\szero)}\left(\frac{\partial \omega_{j}(\bf q)}{\partial V}\right)_{\sV_{\szero}},
\label{Eq:mode-gamma-qha}
\end{eqnarray}

\par
For metals and small band gap materials, the electronic contribution can be considerable and is included in the thermodynamic definitions.
After including the electronic contribution, $\alpha_\sV$ and $C_{\sP}$ are re-formulated as~\cite{Hermet_JPCC_2014,Solyom_Springer_2007}:

\begin{equation}\label{Eq:alpha_el}
\begin{split}
\alpha_{\sV,{\sphelec}}&=\alpha_{\sV}+\alpha_{\sV,{\selectr}} = \frac{C_{\sV}{\bar \gamma}}{B V_{{\sEQ}}} + \frac{2}{3B V_{\sEQ}}C_{{\sV},{\selectr}},\\[5pt]
C_{{\sP},{\sphelec}}  &= C_{\sV}+ V_{\sEQ}TB\alpha_{\sV}^2 + C_{{\sV},{\selectr}},\\[5pt]
C_{{\sV},{\selectr}}&=T\left(\frac{\partial S_{\selectr}}{\partial T}\right)_\sV.
\end{split}
\end{equation}

In addition to the basic quasi-harmonic thermodynamic properties, the enthalpy $\left( H \right)$ of a structure at $P$=0~\cite{Duong_jap_2011,Arroyave_ACTAMAT_2005} is:
\setlength{\abovedisplayskip}{3pt}
\setlength{\belowdisplayskip}{3pt}
\begin{equation}\label{Eq:Heqs}
H=F\left(T\right)+TS=F\left(T\right)-T\frac{\partial{F\left(T\right)}}{\partial T}.
\end{equation}


\begin{figure*}
\centering
\includegraphics*[width=0.92\textwidth]{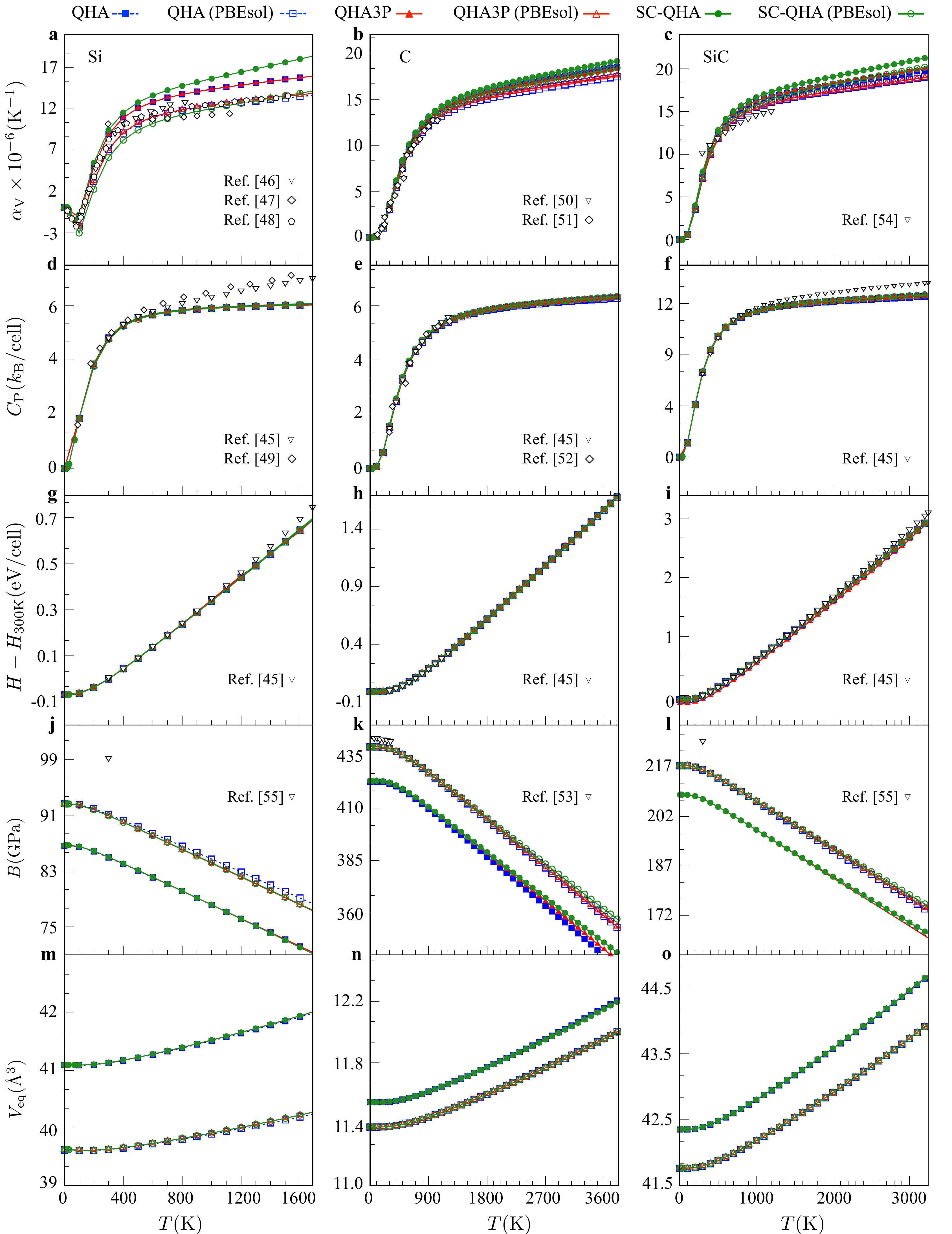}
\vspace{-2mm}
\caption{\small The thermodynamic properties of Si ({\textbf {a,d,g,j,m}}), C ({\textbf {b,e,h,k,n}}), and SiC ({\textbf {c,f,i,l,o}}) are presented up to their melting temperatures ($T_{\protect \scalebox{0.6}{m}}$): $1687$, $3823$ and $3000$~K~\cite{Barin_1993}, respectively.
The results are obtained with \QHAPPP, \QHA\ and \SCQHA\ using the \SPBE\ and \SPBESOL\ \SXC\ functionals.
Experimental results from different sources~\cite{Gibbons_PR_1958,Yim_JAP_1978,Okada_JAP_1984,Chase_AIP_1998,Barin_1993,Slack_JAP_1975,Skinner_AM_1957,Victor_JCP_1962,McSkimin_JAP_1972,Li_JAP_1986,Csonka_PRB_2009} are indicated by inverted triangles, diamonds, and pentagons.}
\label{Fig:fig2}
\end{figure*}

\vspace{5mm}
\noindent\textbf{\AFLOW\ implementation of \SCQHA.}
\SCQHA\ has also been implemented within \AFLOW\ following the description of Ref.~\cite{Huang_cms_2016}.
Similar to \QHA, \SCQHA\ calculates $E_\szero$ at ${\sim}10$ different volumes, but requires only three phonon calculations at different cell volumes (Figure~\ref{Fig:fig1}{\bf c})~\cite{Huang_cms_2016}.
It computes the temperature-dependent unit-cell volume by optimizing the total pressure (external, electronic, and phononic pressures), 

\begin{equation}\label{Eq:V-exp}
V=\left[\left(P-P_\sEE\right)\right]^{-1}\times \frac{1}{N_q} \sum_{{\bf q},j} U_{j}({\bf q})\gamma_{j}({\bf q}),
\end{equation}

\noindent
where $P$ is external pressure, $P_\sEE\left(=-{\rm d} E_\szero\left(V\right)/{\rm d} V\right)$ is the electronic pressure, 
$U_{j}({\bf q})$ is the mode vibrational internal energy and $\gamma_{j}(\bf q)$ is the mode \Gruneisen\ parameter at (${\bf q}, j$).
The volume dependent $\omega_{j}({\bf q})$ and $\gamma_{j}({\bf q})$ are extrapolated to other volumes using a Taylor expansion:

\begin{equation}
\begin{split}
\omega_{j}\left({\bf q}, V\right)&=\omega_{j}\left({\bf q}, V_\szero\right)+\left(\frac{\partial \omega_{j}({\bf q})}{\partial V}\right)_{\sV_{\szero}}(\Delta V)\\
      &+\frac{1}{2}\left(\frac{\partial^2\omega_{j}({\bf q})}{\partial V^2}\right)_{\sV_{\szero}}\left(\Delta V\right)^2,
\end{split}
\label{Eq:mode-omega}
\end{equation}

\begin{equation}
\gamma_{j}\left({\bf q}, V\right)=-\frac{V}{\omega_{j}({\bf q})}\left[\left(\frac{\partial \omega_{j}({\bf q})}{\partial V}\right)_{\sV_{\szero}}+\left(\frac{\partial^2\omega_{j}({\bf q})}{\partial V^2}\right)_{\sV_{\szero}}\Delta V\right],
\label{Eq:mode-gamma}
\end{equation}

\noindent
where  $\Delta V{=}V{-}V_\szero$.
Computing the second order derivative of $\omega_{j}({\bf q})$ requires the calculation of phonon spectra at three different volumes.
Due to its numerical accuracy, the central difference algorithm is used to calculate the derivative with respect to volume. 
$U_{j}({\bf q})$, the mode vibrational internal energy, is defined as 

\begin{eqnarray}
U_{j}({\bf q})&=&\left[ \left({\EXP\left(\frac{\hbar\omega_{j}({\bf q})}{k_{{\sB}}T}\right)-1} \right)^{-1} + \frac{1}{2}\right]  \hbar \omega_{j}({\bf q}). 
\end{eqnarray}

The procedure to self-consistently optimize the volume at zero external pressure ($P{=}0$) and finite $T$ is as follows~\cite{Huang_cms_2016} (Figure~\ref{Fig:fig1}{\bf c}):
\begin{enumerate}[label=(\arabic*)]
  \item First, ${\sim} 10$ $E_\szero\left(V\right)$ values are fitted to the \EOS, enabling the analytical calculation of $P_\sEE$ at any new volume.
  \item $\left(\partial \omega_{j}({\bf q})/\partial V\right)_{\sV_{\szero}}$ and $\left(\partial^2 \omega_{j}({\bf q})/\partial V^2\right)_{\sV_{\szero}}$ are calculated from the three phonon spectra, where $\omega_{j}({\bf q})$ and $\gamma_{j}({\bf q})$ are initialized to their values at $V_\szero$.
  \item To compute the equilibrium volume at $T$, $V$ is initialized to a value $0.2\%$ larger than $V_\szero$ and the following loop is iterated:
  \begin{enumerate}[label=(\roman*)]
    \item Calculate $P_\sEE$ using the \EOS, and $\sum_{{\bf q},j}U_{j}({\bf q})\gamma_{j}({\bf q})$ using $\omega_{j}({\bf q})$ and $\gamma_{j}({\bf q})$.
    \item Update the value of $V$ using Eq.~(\ref{Eq:V-exp}), and update $\omega_{j}({\bf q})$ and $\gamma_{j}({\bf q})$ using Eq.~(\ref{Eq:mode-omega} - \ref{Eq:mode-gamma}), respectively.
    \item If $V$ is not converged to within an acceptable threshold ({\it e.g}., $10^{-6}$), then loop over steps (i) and (ii). 
    \item Calculate other thermodynamic properties using the converged values of  $\omega_{j}({\bf q})$ and $\gamma_{j}({\bf q})$.
    \item $V$, $\omega_{j}({\bf q})$, and $\gamma_{j}({\bf q})$ values at a given $T$ are used as the initial values for the next $T$ characterization.
 \end{enumerate}
\end{enumerate}

In \SCQHA, at low temperatures ({\it e.g}., $0.1$~K), $V_\sEQ$ is calculated self-consistently.
At higher temperatures, $V_\sEQ$ is extrapolated as $V_{\sEQ}\left(T+\Delta T\right) \simeq \left(1+\alpha_\sV \Delta T\right)V_\sEQ$, as described in Ref.~\cite{Huang_cms_2016}, in order to avoid self-consistent volume calculations for each $T$.
$V_\sEQ$ is equal to the self-consistently converged $V$ at a given $T$. 
Similarly, all of the properties calculated at $V_\sEQ$ are the equilibrium properties at $T$. 

While $C_\sV$, $C_\sP$, and $\alpha_\sV$ are the same as for \QHA, $B$ and $\gamma_{j}({\bf q})$ are computed differently in \SCQHA.
Here:

\begin{eqnarray}
B & = & V_{{\sEQ}}\left(\frac{{\rm d}^2 E_\szero} {{\rm d} V^2}\right)_{V_{{\sEQ}}} + B_{\gamma}+B_{\Delta \gamma}+P_{\gamma},
\label{Eq:B-scqha}
\end{eqnarray}

\noindent
where $B_{\gamma}$ and $B_{\Delta \gamma}$ are the bulk modulus contributions due to phonons, while $P_{\gamma}$ represents the  bulk modulus contribution due to external pressure.
The mathematical expressions for these variables are defined in Ref.~\cite{Huang_cms_2016},
and $\gamma_{j}({\bf q})$ is computed using Eq.~(\ref{Eq:mode-gamma}).
It is also important to note that the value of $C_\sV$ at given $T$ is computed with the $\omega_{j}({\bf q})$ values instead of $\omega_{j}({\bf q},V_\szero)$,
where $\omega_{j}({\bf q})$ is the volume (and thus temperature) dependent frequency.

\par
The temperature dependent electronic energy contributions can be added to \SCQHA\ by establishing the relationship between the electronic
eigenvalues and $V$ (similar to the relationship between phonon eigenvalues and volume (Eq.~\ref{Eq:V-exp})).
$F_\selectr$ is not included in the version of \SCQHA\ implemented in \AFLOW.
Derivations and a more detailed description of \SCQHA\ can be found in Ref.~\cite{Huang_cms_2016}.
The version of \SCQHA\ implemented in \AFLOW\ is equivalent to the 2nd-\SCQHA.

\begin{figure*}
\centering
\includegraphics*[width=0.92\textwidth]{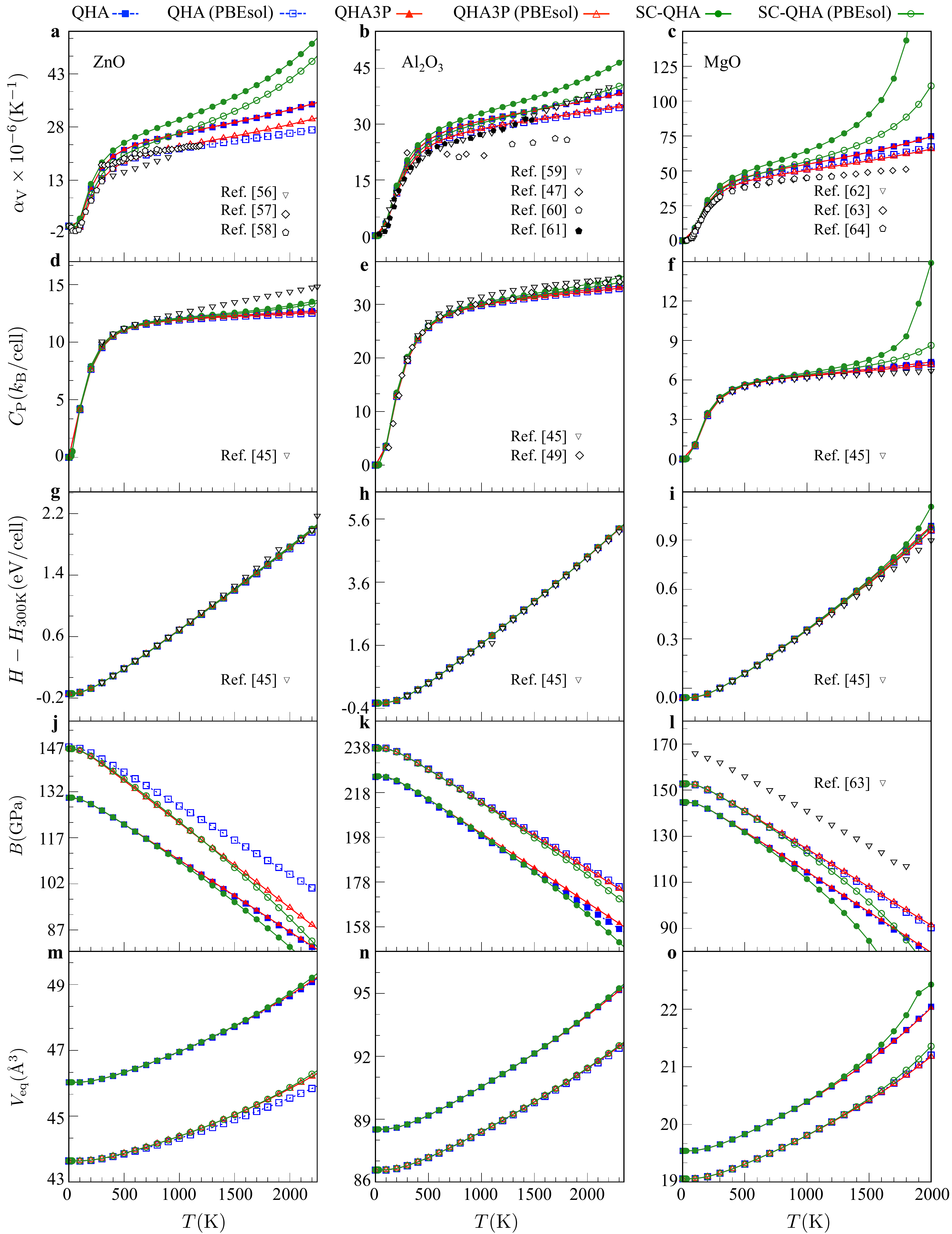}
\vspace{-2mm}
\caption{\small The thermodynamic properties of ZnO ({\textbf {a,d,g,j,m}}), Al$_2$O$_3$ ({\textbf {b,e,h,k,n}}), and MgO ({\textbf {c,f,i,l,o}}) as a function of temperature.
The results of ZnO and  Al$_2$O$_3$ are presented up to $T_{\protect \scalebox{0.6}{m}}$, while MgO  is shown up to $(2/3)T_{\rm m}$.
The $T_{\protect \scalebox{0.6}{m}}$ values of ZnO, Al$_2$O$_3$ and MgO are $2248$, $2345$ and $3125$~K~\cite{Barin_1993}, respectively.
The results are obtained with \QHAPPP, \QHA\ and \SCQHA\ using the \SPBE\ and \SPBESOL\ \SXC\ functionals.
Experimental results from different sources~\cite{Khan_ACTACRIST_1986,Touloukian_NY_1977,Ibach_PSS_1972,Kondo_JPNJAP_2008,Yim_JAP_1978,Munro_JACERS_1997,Wachtman_JACERS_1962,Madelung_Springer_1999,Anderson_RGP_1992,White_JAP_1996,Barin_1993} are indicated by inverted triangles, diamonds, pentagons, and filled pentagons.}
\label{Fig:fig3}
\end{figure*}


\vspace{5mm}
\noindent\textbf{\QHAPPP\ Methodology.}
The \QHAPPP\ method requires only three phonon calculations along with the ${\sim}10$ $E_\szero$ energies. 
The Taylor expansion (Eq.~\ref{Eq:mode-omega}) introduced in \SCQHA~\cite{Huang_cms_2016}, is used to extrapolate the phonon spectra to the remaining volumes.
The following steps are the same as \QHA: fit to the \EOS, minimize $F$ with respect to $V$, and calculate the thermodynamic properties (Figure~\ref{Fig:fig1}{\bf a}).

Although the same technique is used to extrapolate the phonon frequencies (Eq.~(\ref{Eq:mode-omega})) in both \QHAPPP\ and \SCQHA, the definitions of some thermodynamic properties and the method of computing them are different.
For example, the mode \Gruneisen\ in \QHAPPP\ is calculated using Eq.~(\ref{Eq:mode-gamma-qha}), whereas in \SCQHA\ it is obtained using Eq.~(\ref{Eq:mode-gamma}). 
In \SCQHA, $\gamma_{j}({\bf q})$ is temperature dependent, whereas it is temperature independent in \QHAPPP\ and \QHA. 


\noindent
{\it Implications of T dependent $\gamma_{j}({\bf q})$ in \SCQHA.}
The main contribution to Eq.~(\ref{Eq:mode-gamma}) comes from $V_{{\sEQ}}$, $\omega_{j}({\bf q})$, and $\left(\partial \omega_{j}({\bf q})/\partial V\right)_{\sV_{\szero}}$.
The second term is neglected due to the small size of $\left(\partial^2\omega_{j}({\bf q})/\partial V^2\right)_{\sV_{\szero}}$.

\begin{itemize}
  \item In \SCQHA, $\gamma_{j}({\bf q})$ is temperature dependent due to the prefactor $V_\sEQ\left(T\right)/\omega_{j}\left({\bf q}, V_{{\sEQ}}\left(T\right)\right)$ (Eq.~(\ref{Eq:mode-gamma})).
  \item  If $V_{{\sEQ}}$ changes positively with $T$ and $\Big{(}\partial \omega_{j}({\bf q})/\partial V\Big{)}_{\sV_{\szero}}<0$, then $\gamma_{j}({\bf q})$ increases with $T$ (Eq.~(\ref{Eq:mode-gamma})).
  \item $\omega_{j}\left({\bf q}, V_{{\sEQ}}\left(T\right)\right)$ should decrease with increasing $T$ if the above condition is valid (Eq.~(\ref{Eq:mode-omega})), amplifying the $T$ dependence of $\gamma_{j}({\bf q})$.
  \item If $\gamma_{j}({\bf q})$ increases for the majority of (${\bf q},j$) in the Brillouin zone, then $\bar{\gamma}$ will be overestimated by \SCQHA\ in comparison to \QHAPPP\ and \QHA.
  \item ${\left(\partial \omega_{j}({\bf q})/\partial V \right)}_{\sV_{\szero}}$ is negative when $\omega_{j}({\bf q})$ decreases with $T$. The behavior of ${\left(\partial \omega_{j}({\bf q})/\partial V\right)}_{\sV_{\szero}}$ is described by the following expressions~\cite{vanRoekeghem_PRB_2016}:
\end{itemize}

\begin{equation}\label{Eq:dvdt}
\begin{split}
\left.\frac{{\rm d}\omega_{j}({\bf q})} {{\rm d}T}\right|_{\sP} &=  \left(\frac{\partial \omega_{j}({\bf q})}{\partial V}\right)_{P} \left.\frac{\partial V}{ \partial T}\right|_{\sP} \\[5pt]
\Rightarrow \left(\frac{\partial \omega_{j}({\bf q})}{\partial V}\right)_{\sP} &=\frac{1}{V \alpha_\sV}\left.\frac{{\rm d}\omega_{j}({\bf q)}} {{\rm d}T}\right|_{\sP}.
\end{split}
\end{equation}

Thus, \SCQHA\ always produces larger $\bar{\gamma}$ values than \QHAPPP\ and \QHA\ for materials where $\omega_{j}({\bf q})$ decreases with $T$ for the majority of (${\bf q},j$), and that have positive thermal expansion. 
This applies to positive thermal expansion materials (all of the materials in this study).
The effect of this on $\alpha_\sV$ is discussed in the next section.

\vspace{5mm}
\noindent\textbf{The root-mean-square relative deviation.}
The root-mean-square relative deviation (\SRMSRD)

\begin{equation}\label{Eq:RMSrD}
 \chi\left({\small X},{\small Y}\right) = \sqrt{ \frac{\sum_{i}^{N}\left(\frac{X_{i}-Y_{i}}{X_{i}}\right)^2}{N-1} }
\end{equation}

\noindent
is used to quantitatively compare the calculated thermodynamic properties between the various methods, and to validate the model against experiments.
Here $ \chi\left({\small X},{\small Y}\right)$ represents the \SRMSRD\ between $N$ data points obtained from methods $\small X$ and $\small Y$.
Small values of $ \chi\left({\small X},{\small Y}\right)$ indicate that $\small X$ and $\small Y$ produce statistically similar results.

\vspace{5mm}
\noindent\textbf{Geometry optimization.}
All structures are fully relaxed using the high-throughput framework \AFLOW, and density functional theory package \VASP\ ~\cite{kresse_vasp}. 
Optimizations are performed following the \AFLOW\ standard~\cite{curtarolo:art104}. 
The \underline{p}rojector \underline{a}ugmented \underline{w}ave ({\small PAW}) pseudopotentials~\cite{PAW} and the \SXC\ functionals proposed by 
\underline{P}erdew-\underline{B}urke-\underline{E}rnzerhof (\SPBE)~\cite{PBE} are used for all calculations, unless otherwise stated. 
To study functional effects and compare with previous \SCQHA\ results~\cite{Huang_cms_2016},
calculations using the \SPBESOL\ functional~\cite{Perdew_PRL_2008} are also carried out.
A high energy cutoff (40$\%$ larger than the maximum recommended 
cutoff among all the species) and a  \textbf{k}-point mesh of 8000 \textbf{k}-points per reciprocal atom are used to ensure the accuracy of the results.
Primitive cells are fully relaxed until the energy difference between two consecutive ionic 
steps is smaller than $10^{-9}$ eV and forces on each atom are below $10^{-8}$ eV/$\rm{\AA}$.

\vspace{5mm}
\noindent\textbf{Phonon calculations.}
Phonon calculations are carried out using the Automatic Phonon Library~\cite{curtarolo:art119,curtarolo:art125}, as implemented in \AFLOW, using \VASP\ to obtain the interatomic force constants via the finite-displacements approach. 
The magnitude of this displacement is chosen as 0.015$\rm \AA$. 
Supercell size and the number of atoms in the supercell (supercell atoms) along with space group number (sg \#) of each example~\cite{curtarolo:art135} are listed in Table~\ref{tbl:scsize}. 
Non-analytical contributions to the dynamical matrix are also included using the formulation developed by Wang \emph{et al.}~\cite{Wang2010}. 
Frequencies and other related phonon properties are calculated on a 31$\times$31$\times$31 mesh in the Brillouin zone: sufficient to converge the vibrational density of states and thus the corresponding thermodynamic properties.
The phonon density of states is calculated using the linear interpolation tetrahedron technique available in \AFLOW. 

The \QHA\ calculations are performed on 10 equally spaced volumes ranging from ${-}3 \%$ to $6 \%$ uniform strain with $1 \%$ increments from the respective equilibrium structures of the crystal.
More expanded volumes are used since most materials have positive thermal expansion.
Both the \SCQHA\ and the \QHAPPP\ calculations are performed using $\pm 3\%$ expanded and compressed volumes.
All calculations are performed without external pressure ($P$=0), and all volume distortions are isotropic.

\begin{table}
\caption{\small Compound names with ICSD number, supercell size, supercell atoms, lattice type, and sg \#. More detailed information is available in Table~\ref{tbl:scsizeL} Appendix.}
\centering
\def\arraystretch{1.5}
\begin{tabular}{ | c | c | C{1.5cm} | C{1.5cm} | C{1.2cm} | c |}
\hline
  compound  & ICSD   & supercell size    & supercell atoms & lattice type & sg \#\\
\hline
 Si   & ~76268 & 5$\times$5$\times$5  &  250   &   fcc  & 227 \\
 C (Diamond)   & ~28857 & 4$\times$4$\times$4  &  128   &   fcc  & 227 \\
 SiC     & 618777 & 4$\times$4$\times$3  &  192   &   hex  & 186 \\
 Al$_2$O$_3$   & ~89664 & 2$\times$2$\times$2  &  ~80   &   rhl  & 167 \\
 MgO     & 159372 & 4$\times$4$\times$4  &  128   &   fcc  & 225 \\
 ZnO     & 182356 & 4$\times$4$\times$3  &  192   &   hex  & 186 \\
 AlNi    & 602150 & 4$\times$4$\times$4  &  128   &   cub  & 221 \\ 
 NiTiSn  & 174568 & 3$\times$3$\times$3  &  ~81   &   fcc  & 216 \\ 
 Ti$_2$AlN  & 157766 & 4$\times$4$\times$1  &  128   &   hex  & 194 \\
\hline
\end{tabular}
\label{tbl:scsize}
\end{table}

%
\begin{table*}[h!]
\caption{\small \SRMSRD\ for $\alpha_{\protect \scalebox{0.6}{V}}$ for all non-metallic materials.
$\chi$ for $\alpha_{\protect \scalebox{0.6}{V}}$ are calculated with respect to the experiments in: Ref.~\cite{Okada_JAP_1984} (Si), Ref.~\cite{Slack_JAP_1975} (C), Ref.~\cite{Li_JAP_1986} (SiC), Ref.~\cite{Touloukian_NY_1977} (ZnO), Ref.~\cite{Kondo_JPNJAP_2008} (Al$_2$O$_3$), and Ref.~\cite{Anderson_RGP_1992} (MgO). Units: \SRMSRD\ in \%.}
\def\arraystretch{1.5}%
\centering
\begin{tabular}{  | c | c | c | c | c | c | c | c | c | }
\hline
 & {$\chi({\protect \scalebox{0.6}{QHA3P}}, {\protect \scalebox{0.6}{QHA}})$ \SC{\%}} 
 & {$\chi({\protect \scalebox{0.6}{SC-QHA}},{\protect \scalebox{0.6}{QHA}})$ \SC{\%}} 
 & {$\chi({\protect \scalebox{0.6}{QHA3P}}, {\protect \scalebox{0.6}{expt}})$ \SC{\%}}
 & {$\chi({\protect \scalebox{0.6}{SC-QHA}},{\protect \scalebox{0.6}{expt}})$ \SC{\%}} \\ 
     &  \SC{($\alpha_\sV$)}&  \SC{($\alpha_\sV$)}& \SC{($\alpha_\sV$)} & \SC{($\alpha_\sV$)} \\
\hline
 compound  & ~~\SPBE~~~\SPBESOL\ & ~~\SPBE~~~\SPBESOL\ &~\SPBE~~~\SPBESOL\ & \SPBE~~~\SPBESOL\ \\ 
\hline
   Si   & 0.1~~~~~~1.3~~   &   ~9.2~~~~~~7.5~~   &  15.7~~~~~~8.4~~  &  25.8~~~~~19.6~~  \\
   C    & 0.8~~~~~~1.0~~   &   ~2.8~~~~~~4.5~~   &  14.0~~~~~12.1~~  &  16.8~~~~~12.4~~  \\
   SiC     & 0.7~~~~~~0.4~~   &   ~4.7~~~~~~3.7~~   &  13.5~~~~~12.7~~  &  12.6~~~~~11.1~~  \\
   ZnO     & 0.4~~~~~~5.9~~   &  ~27.0~~~~~37.7~~   &  16.7~~~~~~7.3~~  &  31.7~~~~~16.2~~  \\
   Al$_2$O$_3$   & 0.3~~~~~~0.8~~   &  ~11.1~~~~~~9.5~~   &  12.4~~~~~~8.1~~  &  18.7~~~~~~9.1~~  \\
   MgO     & 0.1~~~~~~1.8~~   &   27.5~~~~ 14.5~~   &  25.4~~~~~14.4~~  &  75.0~~~~~38.0~~  \\
\hline
\end{tabular}
\label{chi-table}
\end{table*}
%
\begin{table*}[h!]
\caption{\small \SRMSRD\ for $\alpha_{\protect \scalebox{0.6}{V}}$, $C_{\protect \scalebox{0.6}{P}}$, and $H$ for metallic materials and a small band gap compound.
$\chi$ for $\alpha_{\protect \scalebox{0.6}{V}}$ are computed with respect to the experiment in Refs:~\cite{Wang_ACTAMAT_2004} (AlNi), Ref.~\cite{Hermet_JPCC_2014} (NiTiSn), and Ref.~\cite{Drulis_JAC_2007} (Ti$_2$AlN). Units: \SRMSRD\ in \%.}
\centering
\def\arraystretch{1.5}%
\begin{tabular}{ | c | c | c | c | c | c | c | }
\hline
   Compound
 & {$\chi({\protect \scalebox{0.6}{QHA3P}},{\protect \scalebox{0.6}{expt}})$ \SC{\%} }
 & {$\chi({\protect \scalebox{0.6}{QHA3P}},{\protect \scalebox{0.6}{expt}})$ \SC{\%} }
 & {$\chi({\protect \scalebox{0.6}{QHA3P}},{\protect \scalebox{0.6}{expt}})$ \SC{\%} }
 & {$\chi({\protect \scalebox{0.6}{QHA3P}},{\protect \scalebox{0.6}{expt}})$ \SC{\%} }
 & {$\chi({\protect \scalebox{0.6}{QHA3P}},{\protect \scalebox{0.6}{expt}})$ \SC{\%} }
 & {$\chi({\protect \scalebox{0.6}{QHA3P}},{\protect \scalebox{0.6}{expt}})$ \SC{\%} } \\
     & \SC{($\alpha_{\sV,\sphelec}$)} & \SC{($\alpha_\sV$)} & \SC{($C_{\sP,\sphelec}$)} & \SC{($C_{\sP}$)} & \SC{($H_{\sphelec}$)} & \SC{($H$)} \\
\hline
{\small AlNi}    &  5.2 & 5.1 & 2.6 & 3.2 & 5.6 & 1.9 \\ 
{\small NiTiSn}  &  6.4 & 6.2 & 0.0 & 0.0 & n/a & n/a \\
{\small Ti$_2$AlN}  &  4.9 & 5.7 & 0.0 & 0.0 & n/a & n/a \\
\hline
\end{tabular}
\label{chi-table2}
\end{table*}
\begin{figure*}[h!]
\centering
\includegraphics*[width=0.8\textwidth]{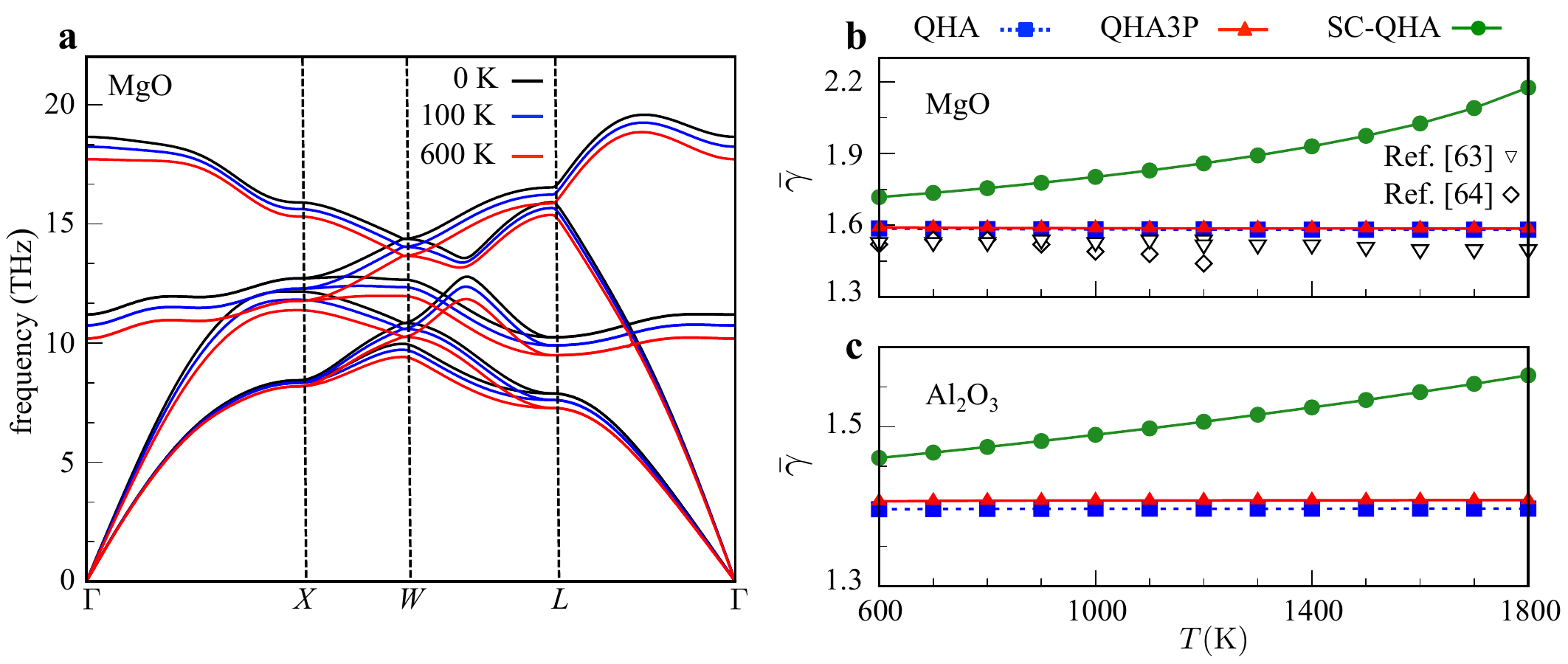}
\vspace{-3mm}
\caption{\small ({\textbf a}) The phonon dispersion curves of MgO at three different temperatures are computed using \SCQHA\ equilibrium volumes and Eq.~(\ref{Eq:mode-omega}).
({\textbf{b, c}}) The $\bar{\gamma}$ values as a function of temperature for MgO and Al$_2$O$_3$ using \QHAPPP, \QHA\, and \SCQHA.
The experimental \Gruneisen\ parameters of MgO are taken from Ref.~\cite{White_JAP_1996,Anderson_RGP_1992}.}
\label{Fig:fig4}
\end{figure*}
%
\begin{figure*}
\centering
\includegraphics*[width=0.90\textwidth]{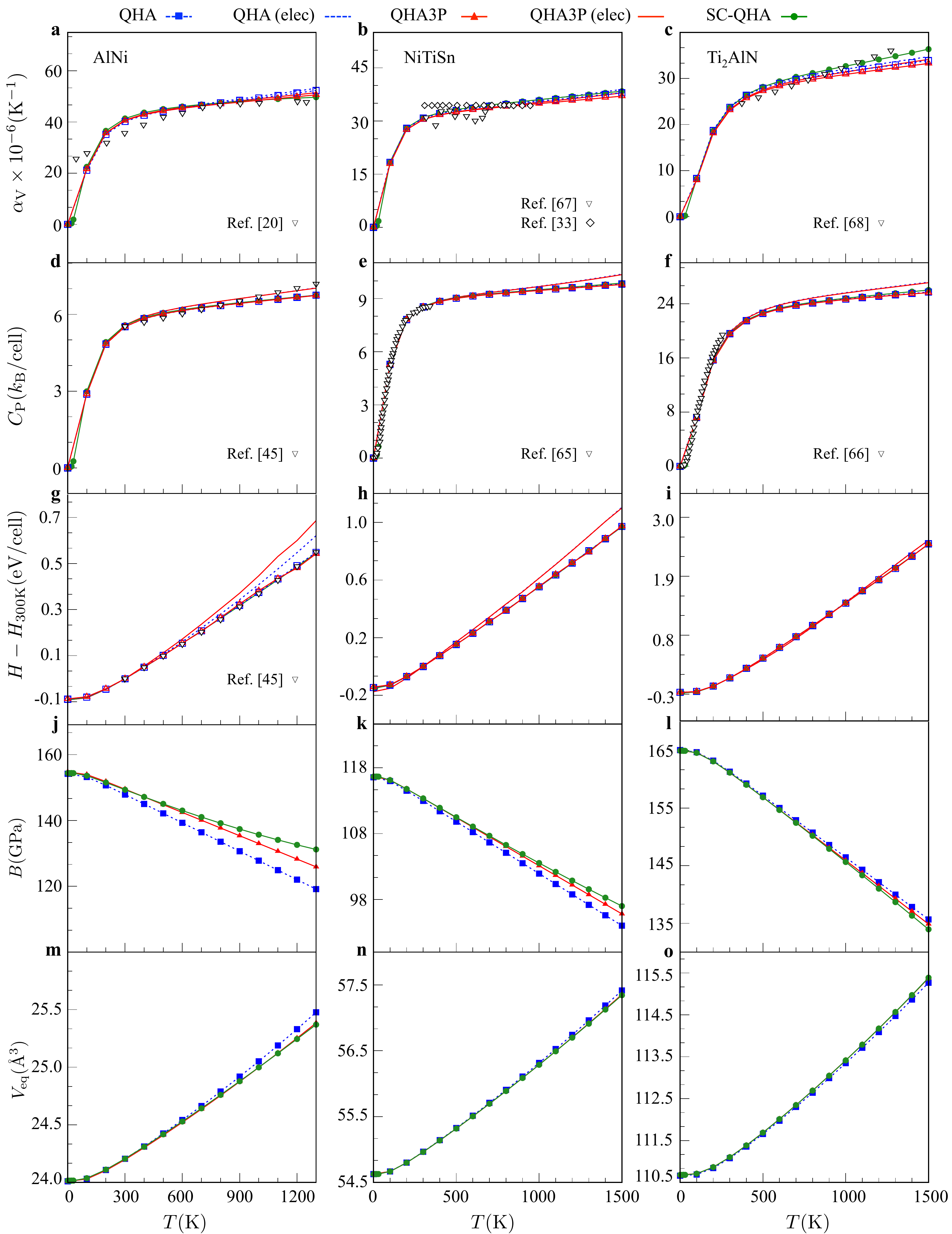}
\vspace{-5mm}
\caption{\small The thermodynamic properties of AlNi ({\textbf {a,d,g,j,m}}), NiTiSn ({\textbf {b,e,h,k,n}}), and Ti$_2$AlN ({\textbf {c,f,i,l,o}}) are shown up to the 
respective $T_{\protect \scalebox{0.6}{m}}$ of $1300$~K, $1500$~K and $1500$~K~\cite{Wang_ACTAMAT_2004, Jung_JAC_2010, Hermet_JPCC_2014, Lane_JCCERS_2011,Drulis_JAC_2007}, respectively.
\QHAPPP\ results are compared with \QHA\ and experiments. 
Experimental results from different sources~\cite{Wang_ACTAMAT_2004, Jung_JAC_2010, Hermet_JPCC_2014, Lane_JCCERS_2011, Drulis_JAC_2007,Barin_1993}
are indicated by inverted triangles and diamonds.}
\label{Fig:fig5}
\end{figure*}

%
%

\section*{\label{sec:results-discussion}Results and Discussion}
Thermodynamic properties of non-metallic materials from the three approaches are compared, and the same methods are applied to metallic and small band gap materials.

\vspace {5mm}
{\bf Non-metallic compounds.}
%
The thermodynamic properties are illustrated in Figure~\ref{Fig:fig2} for Si, C, and SiC and in Figure~\ref{Fig:fig3} for ZnO, Al$_2$O$_3$, and MgO.
Comparisons of \QHAPPP\ with \QHA, \SCQHA, and experimental data are discussed below with two different \SXC\ functionals.

$\alpha_\sV$ from both \QHAPPP\ and \QHA\ agree well with the available experiments for all tested non-metallic compounds using both \SXC\ functionals.
Conversely, \SCQHA\ overestimates $\alpha_\sV$ for Si, ZnO, Al$_2$O$_3$, and MgO.
Using the \SPBESOL\ \SXC\ functional improves consistency with experimental results for Si, however $\alpha_\sV$ is still larger for ZnO and MgO.
For Al$_2$O$_3$, $\alpha_\sV$ values from different experiments ~\cite{Kondo_JPNJAP_2008,Yim_JAP_1978,Munro_JACERS_1997,Wachtman_JACERS_1962} vastly differ from one other.
Below $1500$~K, experimental results from Ref~\cite{Kondo_JPNJAP_2008,Wachtman_JACERS_1962} are well described by \SPBESOL, whereas above $1500$~K, they are better predicted by \SPBE.
The \SRMSRD\ between $\alpha_\sV$ from different approaches and experiment are provided in Table~\ref{chi-table}. 
The relatively small \SRMSRD\ between \QHAPPP\ and experiments compared to the \SRMSRD\ between \SCQHA\ and experiments indicate that the \QHAPPP\ $\alpha_\sV$ predictions are more reliable.

\par
The larger values of $\alpha_\sV$ from \SCQHA\ are due to the overestimation of $\bar\gamma$.
This is particularly significant for positive $\alpha_\sV$ materials, where $\omega_{j}({\bf q})$ decreases with $T$ for almost all (${\bf q},j$) (\QHAPPP\ Methodology section). 
For example, the $T$ dependent phonon dispersion curves of MgO (Figure~\ref{Fig:fig4}(a)) show that $\omega_{j}({\bf q})$ decreases for all (${\bf q},j$), except for the acoustic branches near $\Gamma$.
The $\bar\gamma$ values for MgO along with  Al$_2$O$_3$ are illustrated in Figure~\ref{Fig:fig4}(b, c), and are validated by experiments.
%
\par
Despite differences in $\alpha_\sV$, there are no significant differences between $C_{\sP}$ and $H$ obtained from the three approaches.
The values are insensitive to the choice of functional, except for MgO (Figure~\ref{Fig:fig2} and \ref{Fig:fig3}(d-i)).
Both $C_{\sP}$ and $H$ match well with the experimental values.
A slight underestimation is observed for $C_{\sP}$ for Si, SiC, and ZnO.
Also, divergence from experiments occurs near $2000$~K for MgO using the \SPBE\ \SXC\ functional in \SCQHA.
The observations are supported by the \SRMSRD\ of $C_{\sP}$ (Table~\ref{chi-tableL}, Appendix).
%
\par
The values of $B$ computed with the three methodologies also match when using the same \SXC\ functional, except for ZnO and MgO (Figure~\ref{Fig:fig2} and~\ref{Fig:fig3}(j-o)).
For ZnO, discrepancies are observed between the \QHAPPP\ and \QHA\ results.
Since the bulk modulus value is related to the curvature of $F(V)$ at $V_\sEQ$, any small change in $F(V)$ leads to a larger change in $B$.
For instance, at $2000$~K the difference in $F$ between \QHAPPP\ and \QHA\ for ZnO is approximately $5$ meV at $6\%$ expanded volume of $V_\szero$ (Figure~\ref{Fig:A7}, Appendix), which significantly affects $B$ (Figure~\ref{Fig:fig3}(j)).
The values of $B$ calculated using \SPBESOL\ are larger than for \SPBE, with similar trends previously reported elsewhere~\cite{Csonka_PRB_2009}, and are closer to experiment.
For MgO, some deviation occurs at high temperature between \SCQHA\ and \QHAPPP\ (and \QHA).
\SRMSRD\ with experiment for $B$ has not been calculated due to limited data availability.
For $B$, the small \SRMSRD\ values between \QHAPPP\ and \QHA\ demonstrate that the methods are consistent with each other.
The \SRMSRD\ values between \SCQHA\ and \QHA\ are high for ZnO and MgO (Table~\ref{chi-tableL}, Appendix).
%
\par
As for $B$, predicted $V_\sEQ$ values are similar for all of the methods when using the same \SXC\ functional.
However, the volume predictions obtained using \SPBESOL\ are smaller than those for \SPBE\ (Figure~\ref{Fig:fig2} and~\ref{Fig:fig3}(m-o)).
%
\vspace {5mm}

{\bf Metallic compounds and narrow band gap materials.}
The thermodynamic properties of the metals AlNi and Ti$_2$AlN and the narrow band gap ($0.12$ eV~\cite{Aliev_PBCM_1990}) half-Heusler NiTiSn are presented in Figure~\ref{Fig:fig5}.
For \QHAPPP\ and \QHA, calculations are performed with and without the electronic contribution.
Results for \SCQHA\ are presented without the electronic contribution, since it is not implemented in \AFLOW\ for this method.
In this section calculations are performed only with the \SPBE\ \SXC\ functional. 
%

Similar to the non-metallic examples, there are no prominent differences between $\alpha_\sV$ obtained using these three methods (Figure~\ref{Fig:fig5}(a-c)).
The $\alpha_\sV$ results show the effectiveness of all methodologies in accurately predicting the experimental thermodynamic properties of AlNi, NiTiSn, and Ti$_2$AlN.
In addition, the contribution of $F_{\selectr}$ to $F$ does not alter the results.

The other thermodynamic properties, $C_{\sP}$, $H$, $B$, and $V_\sEQ$, also match well with each other for these three methods and with experiments, with a few exceptions (Figure~\ref{Fig:fig5}(d-f)).
The $H$ values for AlNi from \QHAPPP\ and \QHA\ do not agree with experiments above 500 K when $F_{\selectr}$ is taken into consideration (Figure~\ref{Fig:fig5}(g, j)).
The $B$ values from the three methods show discrepancies stemming from the difference in the $F(V)$ energies as described in the non-metallic compounds section.
The small \SRMSRD\ between \QHAPPP\ and experiment for $\alpha_\sV$, $C_{\sP}$, and $H$ indicate that the results agree well with experiments (Table~\ref{chi-table2}).
Since the \QHA, \SCQHA, and \QHAPPP\ predictions are similar, \SRMSRD\ values are presented only for \QHAPPP.


\section*{\label{sec:Conclusions}Conclusions}

The quasi-harmonic approximation 3-phonons method is introduced to calculate thermodynamic properties of both non-metallic and metallic compounds.
The efficiency of \QHAPPP\ is tested for a range of materials using two different exchange-correlation functionals, and the calculated thermodynamic quantities are in agreement with both \QHA\ and experimental measurements. 
We also show that \SCQHA\ overestimates the average \Gruneisen\ parameter, as well as $\alpha_\sV$, at high temperatures for some materials, while \QHAPPP\ still performs well. 
This study demonstrates that \QHAPPP\ is an ideal framework for the high-throughput prediction of finite temperature materials properties, combining the accuracy of \QHA\ with the computational efficiency of \SCQHA.
\\~\\


\section*{Acknowledgements}

The authors acknowledge support by 
DOD-ONR (N00014-13-1-0635, N00014-16-1-2368, N00014-15-1-2863, N00014-17-1-2090). 
DH acknowledges support from the Department of Defense through the National Defense Science and Engineering Graduate (NDSEG) Fellowship Program.
CO acknowledges support from the National Science Foundation Graduate Research Fellowship under Grant No. DGF1106401.
SC  acknowledges the Alexander von Humboldt Foundation for financial support.
The consortium \AFLOW.org acknowledges Duke University - Center for Material Genomics - for computational support.

\newpage
\clearpage

\begin{widetext}

\appendix


\section*{Appendix}

\renewcommand{\thefigure}{A\arabic{figure}}
\setcounter{figure}{0}
\setcounter{table}{0}
\renewcommand{\thetable}{A\arabic{table}}

\clearpage

%
%
\begin{table*}
  {\centering
    \caption{\small 
      {\bf List of crystallographic properties.} List of compound names with various crystallographic properties along with \AFLOW\ prototype~\cite{curtarolo:art121,Hicks_arXiv_2018} and \AFLOW\ unique identifier (auid)~\cite{curtarolo:art128}.}
    \def\arraystretch{1.5}%
    \begin{tabular}{ | c | c | C{1.5cm} | C{1.5cm} | C{1.5cm} | c | C{2cm} | c | c |}
      \hline
      compound   & ICSD & supercell size   & supercell atoms & lattice type & sg\#   & {\small DFT} band gap (eV) & prototype & auid \\
      \hline
      Si   & 76268   & 5$\times$5$\times$5 &    250    &   fcc  &   227  &    0.61   & A\_cF8\_227\_a~\cite{A_cF8_227_a}     & \href{http://aflow.org/material.php?id=Si1_ICSD_76268}{aflow:ff211836be789f69}  \\
      C (Diamond)   & 28857   & 4$\times$4$\times$4 &    128    &   fcc  &   227  &    4.11   & A\_cF8\_227\_a~\cite{A_cF8_227_a}     & \href{http://aflow.org/material.php?id=C1_ICSD_28857}{aflow:b438e1a25f9c187d}   \\
      SiC     & 618777  & 4$\times$4$\times$3 &    192    &   hex  &   186  &    2.30   & AB\_hP4\_186\_b\_b~\cite{AB_hP4_186_b_b}    & \href{http://aflow.org/material.php?id=C1Si1_ICSD_618777}{aflow:50.644a24702c7dc}  \\
      ZnO     & 182356  & 4$\times$4$\times$3 &    192    &   hex  &   186  &    1.81   & AB\_hP4\_186\_b\_b~\cite{AB_hP4_186_b_b}    & \href{http://aflow.org/material.php?id=O1Zn1_ICSD_182356}{aflow:f30df164c6192045}  \\
      Al$_2$O$_3$   & 89664   & 2$\times$2$\times$2 &    ~80    &   rhl  &   167  &    5.86   & A2B3\_hR10\_167\_c\_e~\cite{A2B3_hR10_167_c_e} & \href{http://aflow.org/material.php?id=Al2O3_ICSD_89664}{aflow:537e800e0a1b75be}   \\
      MgO     & 159372  & 4$\times$4$\times$4 &    128    &   fcc  &   225  &    4.46   & AB\_cF8\_225\_a\_b~\cite{AB_cF8_225_a_b}    & \href{http://aflow.org/material.php?id=Mg1O1_ICSD_159372}{aflow:a2cc05c200330e16}  \\
      AlNi    & 602150  & 4$\times$4$\times$4 &    128    &   cub  &   221  &    0.00   & AB\_cP2\_221\_b\_a~\cite{AB_cP2_221_b_a}    & \href{http://aflow.org/material.php?id=Al1Ni1_ICSD_602150}{aflow:f727dcd6a292301d} \\
      NiTiSn  & 174568  & 3$\times$3$\times$3 &    ~81    &   fcc  &   216  &    0.17   & ABC\_cF12\_216\_b\_c\_a~\cite{ABC_cF12_216_b_c_a} & \href{http://aflow.org/material.php?id=Ni1Sn1Ti1_ICSD_174568}{aflow:7bed936e9d5a44ca} \\
      Ti$_2$AlN  & 157766  & 4$\times$4$\times$1 &    128    &   hex  &   194  &    0.00   & ABC2\_hP8\_194\_d\_a\_f~\cite{ABC2_hP8_194_d_a_f} & \href{http://aflow.org/material.php?id=Al1N1Ti2_ICSD_157766}{aflow:bdc38ae3ca07e398}  \\
      \hline
    \end{tabular}
    \label{tbl:scsizeL}
  }
\end{table*}
%
%
\begin{table*}
\caption{\small 
{\bf List of \SRMSRD\ values.} \SRMSRD\ for $C_{\protect \scalebox{0.6}{P}}$, and $B$ for all non-metallic materials. 
$\chi$ for $C_{\protect \scalebox{0.6}{P}}$ are calculated with respect to the experiments in: Ref.~\cite{Barin_1993}. 
$\chi$ for $H$ and $V_{\protect \scalebox{0.6}{eq}}$ are similar to $C_{\protect \scalebox{0.6}{P}}$ and $B$, respectively, and are not presented.
Units: \SRMSRD\ in \%.}
\def\arraystretch{1.5}%
\centering
\begin{tabular}{  | c | c | c | c | c |}
\hline
& {$\chi({\protect \scalebox{0.6}{QHA3P}},{\protect \scalebox{0.6}{expt}})$ \SC{\%}}
& {$\chi({\protect \scalebox{0.6}{SC-QHA}},{\protect \scalebox{0.6}{expt}})$ \SC{\%}} 
& {$\chi({\protect \scalebox{0.6}{QHA3P}},{\protect \scalebox{0.6}{QHA}})$  \SC{\%}} 
& {$\chi({\protect \scalebox{0.6}{SC-QHA}},{\protect \scalebox{0.6}{QHA}})$ \SC{\%}} \\
     & \SC{($C_\sP$)}  & \SC{($C_\sP$)}   & \SC{($B$)}   &  \SC{($B$)} \\
\hline
 compound  & ~\SPBE~~~\SPBESOL\ & ~\SPBE~~~\SPBESOL\  & ~\SPBE~~~\SPBESOL\ & ~\SPBE~~~\SPBESOL\ \\
\hline
   Si   & 7.4~~~~~7.6  & ~7.1~~~~~7.7  & 0.5~~~~~0.2~~   &   ~0.2~~~~~0.4~~  \\
   C    & 2.2~~~~~2.2  & ~3.1~~~~~1.8  & 0.5~~~~~0.0~~   &   ~1.0~~~~~0.5~~  \\
   SiC     & 4.8~~~~~5.0  & ~4.4~~~~~4.6  & 0.0~~~~~0.1~~   &   ~0.2~~~~~0.3~~  \\
   ZnO     & 8.5~~~~~9.3  & ~5.9~~~~~7.0  & 0.3~~~~~7.0~~   &   ~3.4~~~~~9.8~~  \\
   Al$_2$O$_3$   & 4.5~~~~~5.3  & ~3.0~~~~~4.1  & 0.8~~~~~0.3~~   &   ~1.1~~~~~1.7~~  \\
   MgO     & 9.0~~~~~3.6  & 37.1~~~~11.5  & 0.5~~~~~0.7~~   &   21.6~~~~~6.6~~  \\
\hline
\end{tabular}
\label{chi-tableL}
\end{table*}

%
%
\begin{figure*}
\centering
\includegraphics*[width=0.49\textwidth]{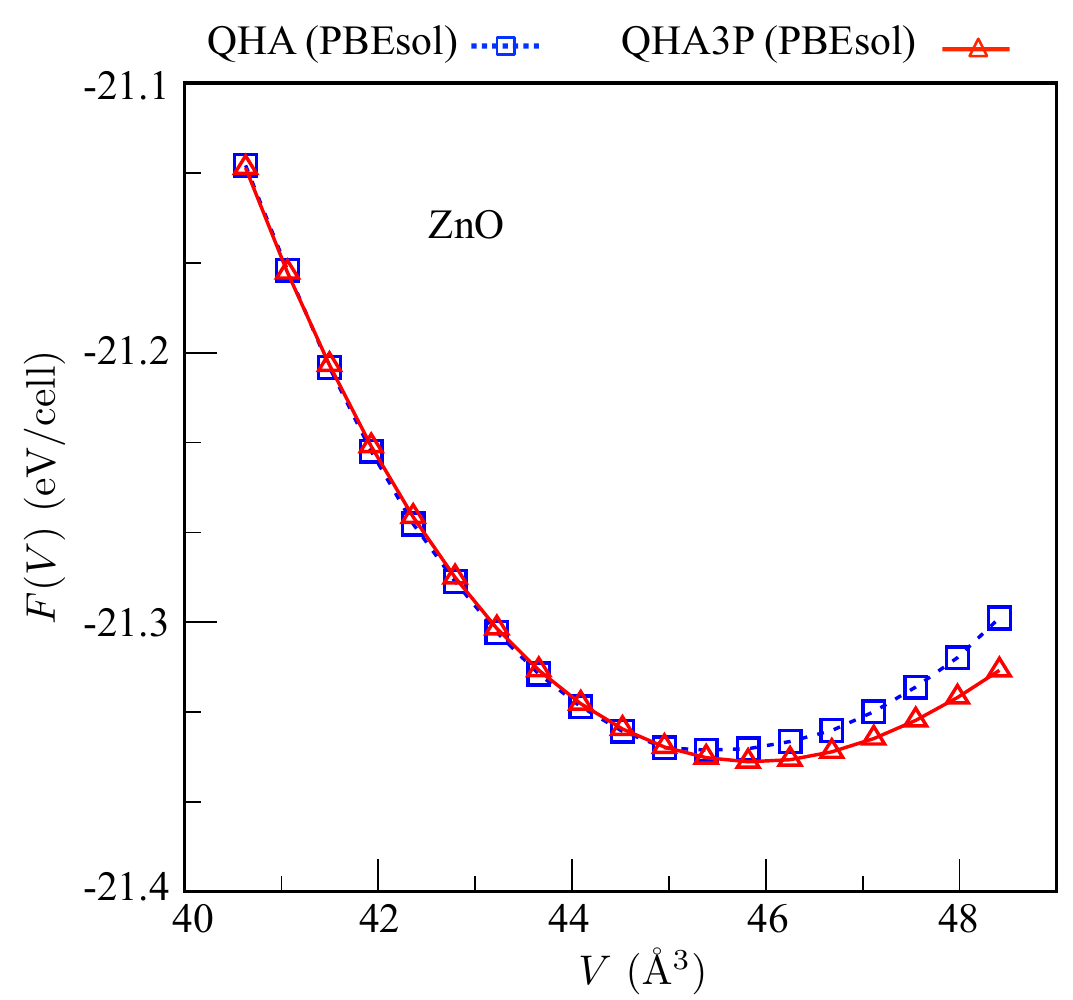}
\vspace{-2mm}
\caption{\small
{\bf $F{-}V$ curve of ZnO.} The $F(V)$ curve of ZnO at $2000$~K. The range of volume is from $-6$ to $12$ percent of $V_{\protect \scalebox{0.6}{0}}$ ($43.22$ \AA$^3$). 
  The difference between the two curves gives $F_{\protect \scalebox{0.6}{vib}}$, since $E_{\protect \scalebox{0.6}{0}}$ is the same for both \QHA\ and \QHAPPP.}
\label{Fig:A7}
\end{figure*}
%
\begin{figure*} 
\includegraphics*[width=0.99\textwidth]{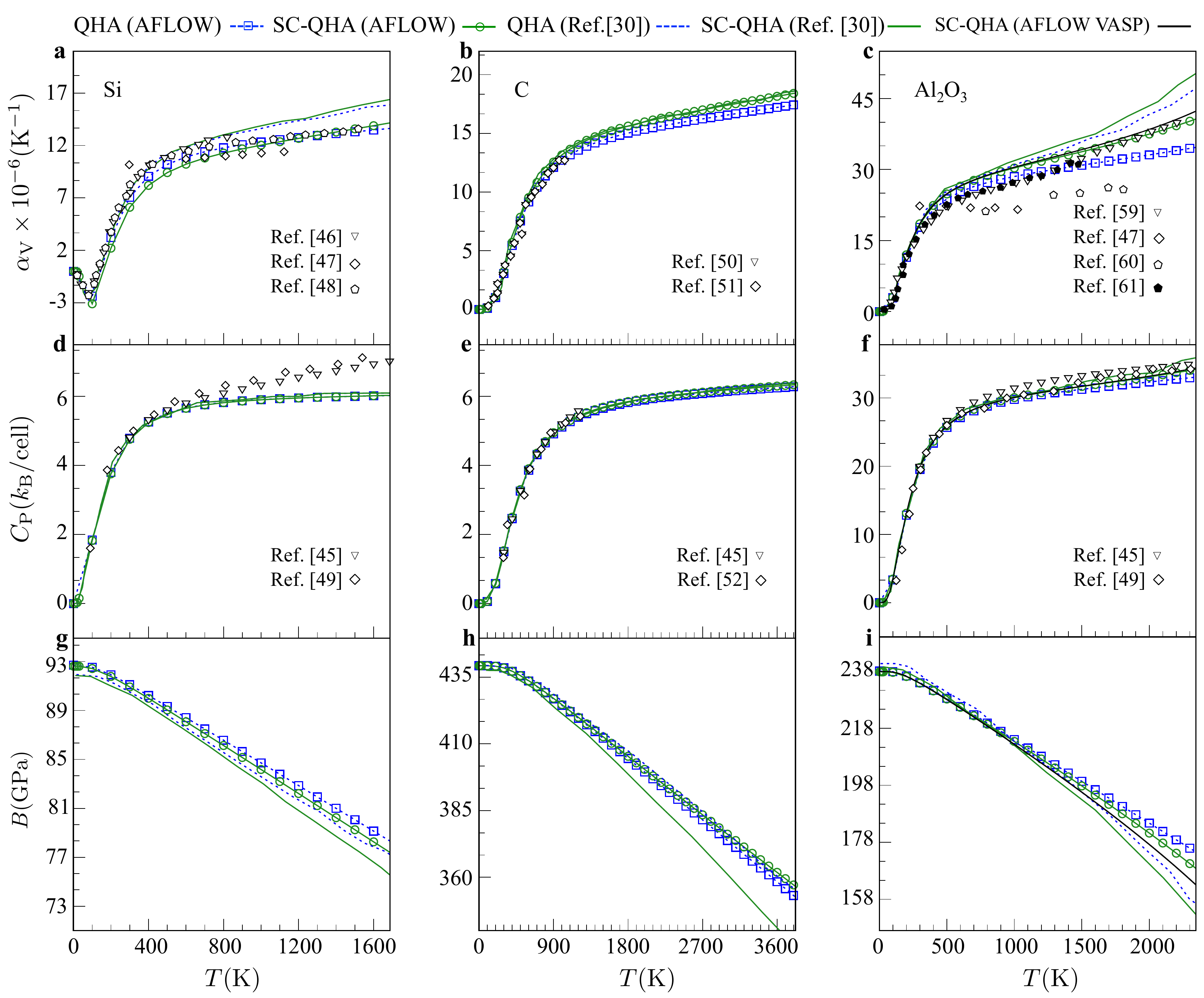}
\vspace{-2mm}
\caption{\small The comparison of \AFLOW\ computed $\alpha_{\protect \scalebox{0.6}{V}}$, $C_{\protect \scalebox{0.6}{P}}$ and $B$ with Ref.~\cite{Huang_cms_2016}.
Experimental results from different sources~\cite{Gibbons_PR_1958,Yim_JAP_1978,Okada_JAP_1984,Kondo_JPNJAP_2008,Skinner_AM_1957,Slack_JAP_1975,Barin_1993,Victor_JCP_1962,Li_JAP_1986,Wachtman_JACERS_1962,Munro_JACERS_1997,Anderson_RGP_1992,Chase_AIP_1998,Barin_1993} are indicated by inverted triangles, diamonds, pentagons and filled pentagons.
The black colored line for Al$_2$O$_3$ represents calculations with the {\small PHONOPY}+\SCQHA\ codes~\cite{Togo_prb_2015,Huang_cms_2016}, performed using the \AFLOW\ standard \VASP\ input parameters.}
\label{Fig:A8}
\end{figure*}

%
{\bf Comparison of \SCQHA\ from \AFLOW\ and  Ref.~\cite{Huang_cms_2016}.} 
The \SCQHA\ method was originally developed and tested in Ref.~\cite{Huang_cms_2016} using the \SPBESOL\ \SXC\ functional.
To check the consistency between the \AFLOW\ implementation of \SCQHA\ and Ref.~\cite{Huang_cms_2016}, the calculated properties are compared for Si, C, and Al$_2$O$_3$.
The differences between the properties computed with \AFLOW\ and  Ref.~\cite{Huang_cms_2016} are  marginal.
However, some discrepancies are observed for the $\alpha_\sV$ values of Si, $B$ values of C, and the $\alpha_\sV$ and $B$ values of Al$_2$O$_3$, which are the presented examples in Ref.~\cite{Huang_cms_2016}.
To investigate this, thermodynamic properties are reproduced for Al$_2$O$_3$ using the original \SCQHA\ and {\small PHONOPY}~\cite{Togo_prb_2015} codes with the \AFLOW\ standard \VASP\ input parameters (Figure~\ref{Fig:A8}, Appendix).
This indicates that the origin of the incompatibility between these two studies is the difference in the \VASP\ input parameters.
While the accuracy is increased by using the \AFLOW\ standard \VASP\ input parameters, the results are still inconsistent with experiments and \QHA.
%


\end{widetext}

\clearpage
\newcommand{\Ozolins}{Ozoli\c{n}\v{s}}


\begin{thebibliography}{10}
\expandafter\ifx\csname urlstyle\endcsname\relax
  \providecommand{\doi}[1]{doi:\discretionary{}{}{}#1}\else
  \providecommand{\doi}{doi:\discretionary{}{}{}\begingroup
  \urlstyle{rm}\Url}\fi
\providecommand{\selectlanguage}[1]{\relax}
\providecommand{\bibAnnoteFile}[1]{%
  \IfFileExists{#1}{\begin{quotation}\noindent\textsc{Key:} #1\\
  \textsc{Annotation:}\ \input{#1}\end{quotation}}{}}
\providecommand{\bibAnnote}[2]{%
  \begin{quotation}\noindent\textsc{Key:} #1\\
  \textsc{Annotation:}\ #2\end{quotation}}

\bibitem{curtarolo:art65}
S.~Curtarolo, W.~Setyawan, G.~L.~W. Hart, M.~Jahn\'{a}tek, R.~V. Chepulskii,
  R.~H. Taylor, S.~Wang, J.~Xue, K.~Yang, O.~Levy, M.~J. Mehl, H.~T. Stokes,
  D.~O. Demchenko, and D.~Morgan, \emph{{AFLOW}: An automatic framework for
  high-throughput materials discovery}, Comput.\ Mater.\ Sci. \textbf{58},
  218--226 (2012).
\input{curtarolo:art65}

\bibitem{curtarolo:art75}
S.~Curtarolo, W.~Setyawan, S.~Wang, J.~Xue, K.~Yang, R.~H. Taylor, L.~J.
  Nelson, G.~L.~W. Hart, S.~Sanvito, M.~{Buongiorno Nardelli}, N.~Mingo, and
  O.~Levy, \emph{{AFLOWLIB.ORG}: A distributed materials properties repository
  from high-throughput {\it ab initio} calculations}, Comput.\ Mater.\ Sci.
  \textbf{58}, 227--235 (2012).
\input{curtarolo:art75}

\bibitem{curtarolo:art92}
R.~H. Taylor, F.~Rose, C.~Toher, O.~Levy, K.~Yang, M.~{Buongiorno Nardelli},
  and S.~Curtarolo, \emph{A {REST}ful {API} for exchanging materials data in
  the {AFLOWLIB}.org consortium}, Comput.\ Mater.\ Sci. \textbf{93}, 178--192
  (2014).
\input{curtarolo:art92}

\bibitem{curtarolo:art127}
A.~R. Supka, T.~E. Lyons, L.~S.~I. Liyanage, P.~{D'{A}mico},
  R.~{Al~Rahal~Al~Orabi}, S.~Mahatara, P.~Gopal, C.~Toher, D.~Ceresoli,
  A.~Calzolari, S.~Curtarolo, M.~{Buongiorno Nardelli}, and M.~Fornari,
  \emph{{\small AFLOW}$\pi$: A minimalist approach to high-throughput {\it ab
  initio} calculations including the generation of tight-binding hamiltonians},
  Comput.\ Mater.\ Sci. \textbf{136}, 76--84 (2017).
\input{curtarolo:art127}

\bibitem{curtarolo:art128}
F.~Rose, C.~Toher, E.~Gossett, C.~Oses, M.~{Buongiorno Nardelli}, M.~Fornari,
  and S.~Curtarolo, \emph{{AFLUX}: The {LUX} materials search {API} for the
  {AFLOW} data repositories}, Comput.\ Mater.\ Sci. \textbf{137}, 362--370
  (2017).
\input{curtarolo:art128}

\bibitem{Yu_SR_2016}
H.~Yu, D.~Duan, H.~Liu, T.~Yang, F.~Tian, K.~Bao, D.~Li, Z.~Zhao, B.~Liu, and
  T.~Cui, \emph{{\it Ab initio} molecular dynamic study of solid-state
  transitions of ammonium nitrate}, Sci.\ Rep. \textbf{6}, 18918 (2016).
\input{Yu_SR_2016}

\bibitem{Kresse_PRB_1993}
G.~Kresse and J.~Hafner, \emph{{\it Ab initio} molecular dynamics for
  open-shell transition metals}, Phys.\ Rev.\ B \textbf{48}, 13115--13118
  (1993).
\input{Kresse_PRB_1993}

\bibitem{Sarnthein_PRB_1996}
J.~Sarnthein, K.~Schwarz, and P.~E. Bl\"{o}chl, \emph{{\it Ab initio}
  molecular-dynamics study of diffusion and defects in solid {{Li}$_{3}${N}}},
  Phys.\ Rev.\ B \textbf{53}, 9084--9091 (1996).
\input{Sarnthein_PRB_1996}

\bibitem{Zhang_APL_2015}
H.~L. Zhang, Y.~F. Han, W.~Zhou, Y.~B. Dai, J.~Wang, and B.~D. Sun,
  \emph{Atomic study on the ordered structure in {Al} melts induced by
  liquid/substrate interface with {Ti} solute}, Appl.\ Phys.\ Lett.
  \textbf{106}, 041606 (2015).
\input{Zhang_APL_2015}

\bibitem{Jesson_APL_2015}
B.~J. Jesson and P.~A. Madden, \emph{Structure and dynamics at the aluminum
  solid-liquid interface: An {\it ab initio} simulation}, J.\ Chem.\ Phys.
  \textbf{113}, 5935--5946 (2000).
\input{Jesson_APL_2015}

\bibitem{Errea_PRL_2013}
I.~Errea, M.~Calandra, and F.~Mauri, \emph{First-Principles Theory of
  Anharmonicity and the Inverse Isotope Effect in Superconducting
  Palladium-Hydride Compounds}, Phys.\ Rev.\ Lett. \textbf{111}, 177002 (2013).
\input{Errea_PRL_2013}

\bibitem{Errea_PRB_2014}
I.~Errea, M.~Calandra, and F.~Mauri, \emph{Anharmonic free energies and phonon
  dispersions from the stochastic self-consistent harmonic approximation:
  Application to platinum and palladium hydrides}, Phys.\ Rev.\ B \textbf{89},
  064302 (2014).
\input{Errea_PRB_2014}

\bibitem{BlancoGIBBS2004}
M.~A. Blanco, E.~Francisco, and V.~Lua{\~n}a, \emph{{GIBBS}:
  isothermal-isobaric thermodynamics of solids from energy curves using a
  quasi-harmonic Debye model}, Comput.\ Phys.\ Commun. \textbf{158}, 57--72
  (2004).
\input{BlancoGIBBS2004}

\bibitem{Poirier_Earth_Interior_2000}
J.-P. Poirier, \emph{Introduction to the Physics of the Earth’s Interior}
  (Cambridge University Press, 2000), 2nd edn.
\input{Poirier_Earth_Interior_2000}

\bibitem{curtarolo:art124}
O.~Isayev, C.~Oses, C.~Toher, E.~Gossett, S.~Curtarolo, and A.~Tropsha,
  \emph{Universal fragment descriptors for predicting electronic properties of
  inorganic crystals}, Nat.\ Commun. \textbf{8}, 15679 (2017).
\input{curtarolo:art124}

\bibitem{curtarolo:art136}
E.~Gossett, C.~Toher, C.~Oses, O.~Isayev, F.~Legrain, F.~Rose, E.~Zurek,
  J.~Carrete, N.~Mingo, A.~Tropsha, and S.~Curtarolo, \emph{{AFLOW-ML}: {A}
  {REST}ful {API} for machine-learning predictions of materials properties},
  Comput.\ Mater.\ Sci. \textbf{152}, 134--145 (2018).
\input{curtarolo:art136}

\bibitem{curtarolo:art96}
C.~Toher, J.~J. Plata, O.~Levy, M.~{de~Jong}, M.~D. Asta, M.~{Buongiorno
  Nardelli}, and S.~Curtarolo, \emph{High-throughput computational screening of
  thermal conductivity, {D}ebye temperature, and {G}r\"{u}neisen parameter
  using a quasiharmonic {D}ebye model}, Phys.\ Rev.\ B \textbf{90}, 174107
  (2014).
\input{curtarolo:art96}

\bibitem{curtarolo:art115}
C.~Toher, C.~Oses, J.~J. Plata, D.~Hicks, F.~Rose, O.~Levy, M.~{de Jong}, M.~D.
  Asta, M.~Fornari, M.~{Buongiorno Nardelli}, and S.~Curtarolo, \emph{Combining
  the {AFLOW} {GIBBS} and Elastic Libraries to efficiently and robustly screen
  thermomechanical properties of solids}, Phys.\ Rev.\ Mater. \textbf{1},
  015401 (2017).
\input{curtarolo:art115}

\bibitem{Liu_Cambridge_2016}
Z.~K. Liu and Y.~Wang, \emph{Computational Thermodynamics of Materials}
  (Cambridge University Press, 2016), 1 edn.
\input{Liu_Cambridge_2016}

\bibitem{Wang_ACTAMAT_2004}
Y.~Wang, Z.-K. Liu, and L.-Q. Chen, \emph{Thermodynamic properties of {Al},
  {Ni}, {Ni}{Al}, and {Ni}$_{3}${Al} from first-principles calculations}, Acta\
  Mater. \textbf{52}, 2665--2671 (2004).
\input{Wang_ACTAMAT_2004}

\bibitem{Duong_jap_2011}
T.~Duong, S.~Gibbons, R.~Kinra, and R.~Arr\'{o}yave, \emph{{\it Ab-initio}
  approach to the electronic, structural, elastic, and finite-temperature
  thermodynamic properties of {Ti}$_{2}{A}{X}$ ({$A$} = {Al} or {Ga} and {$X$}
  = {C} or {N})}, J.\ Appl.\ Phys. \textbf{110}, 093504 (2011).
\input{Duong_jap_2011}

\bibitem{curtarolo:art114}
P.~Nath, J.~J. Plata, D.~Usanmaz, R.~{Al~Rahal~Al~Orabi}, M.~Fornari,
  M.~{Buongiorno Nardelli}, C.~Toher, and S.~Curtarolo, \emph{High-throughput
  prediction of finite-temperature properties using the quasi-harmonic
  approximation}, Comput.\ Mater.\ Sci. \textbf{125}, 82--91 (2016).
\input{curtarolo:art114}

\bibitem{Ziman_Oxford_1960}
J.~M. Ziman, \emph{Electrons and Phonons: The Theory of Transport Phenomena in
  Solids}, Oxford Classic Texts in the Physical Sciences (Oxford University
  Press, 1960).
\input{Ziman_Oxford_1960}

\bibitem{ThermoCrys}
D.~C. Wallace, \emph{Thermodynamics of crystals} (Wiley, 1972).
\input{ThermoCrys}

\bibitem{Dove_LatDynam_1993}
M.~T. Dove, \emph{Introduction to Lattice Dynamics}, Cambridge Topics in
  Mineral Physics and Chemistry (Cambridge University Press, 1993).
\input{Dove_LatDynam_1993}

\bibitem{PhysPhon}
G.~P. Srivastava, \emph{The Physics of Phonons} (CRC Press, Taylor \& Francis,
  1990).
\input{PhysPhon}

\bibitem{BaroniRMP2001}
S.~Baroni, S.~{de Gironcoli}, A.~{Dal Corso}, and P.~Giannozzi, \emph{Phonons
  and related crystal properties from density-functional perturbation theory},
  Rev.\ Mod.\ Phys. \textbf{73}, 515--562 (2001).
\input{BaroniRMP2001}

\bibitem{Axel_RMP}
A.~van~de Walle and G.~Ceder, \emph{The effect of lattice vibrations on
  substitutional alloy thermodynamics}, Rev.\ Mod.\ Phys. \textbf{74}, 11--45
  (2002).
\input{Axel_RMP}

\bibitem{curtarolo:art125}
J.~J. Plata, P.~Nath, D.~Usanmaz, J.~Carrete, C.~Toher, M.~{de Jong}, M.~D.
  Asta, M.~Fornari, M.~{Buongiorno Nardelli}, and S.~Curtarolo, \emph{An
  efficient and accurate framework for calculating lattice thermal conductivity
  of solids: {AFLOW}-{AAPL} {Au}tomatic {A}nharmonic {P}honon {Li}brary}, NPJ\
  Comput.\ Mater. \textbf{3}, 45 (2017).
\input{curtarolo:art125}

\bibitem{Huang_cms_2016}
L.-F. Huang, X.-Z. Lu, E.~Tennessen, and J.~M. Rondinelli, \emph{An efficient
  {\it ab-initio} quasiharmonic approach for the thermodynamics of solids},
  Comput.\ Mater.\ Sci. \textbf{120}, 84--93 (2016).
\input{Huang_cms_2016}

\bibitem{Arroyave_ACTAMAT_2005}
R.~Arroyave, D.~Shin, and Z.-K. Liu, \emph{{\it Ab initio} thermodynamic
  properties of stoichiometric phases in the {Ni}-{Al} system}, Acta\ Mater.
  \textbf{53}, 1809--1819 (2005).
\input{Arroyave_ACTAMAT_2005}

\bibitem{Tohei_JAP_2016}
T.~Tohei, Y.~Watanabe, H.-S. Lee, and Y.~Ikuhara, \emph{First principles
  calculation of thermal expansion coefficients of pure and {Cr} doped
  $\alpha$-alumina crystals}, J.\ Appl.\ Phys. \textbf{120}, 142106 (2016).
\input{Tohei_JAP_2016}

\bibitem{Hermet_JPCC_2014}
P.~Hermet, R.~M. Ayral, E.~Theron, P.~G. Yot, F.~Salles, M.~Tillard, and
  P.~Jund, \emph{Thermal Expansion of {Ni}-{Ti}-{Sn} Heusler and Half-Heusler
  Materials from First-Principles Calculations and Experiments}, J.\ Phys.\
  Chem.\ C \textbf{118}, 22405--22411 (2014).
\input{Hermet_JPCC_2014}

\bibitem{Fu_PRB_1983}
C.-L. Fu and K.-M. Ho, \emph{First-principles calculation of the equilibrium
  ground-state properties of transition metals: Applications to {Nb} and {Mo}},
  Phys.\ Rev.\ B \textbf{28}, 5480--5486 (1983).
\input{Fu_PRB_1983}

\bibitem{Solyom_Springer_2007}
J.~S\'{o}lyom, \emph{Fundamentals of the physics of solids}, vol.~1 (Springer
  Berlin Heidelberg, 2007), 1 edn.
\input{Solyom_Springer_2007}

\bibitem{Barin_1993}
I.~Barin, \emph{Thermochemical Data of pure substances} (VCH, Germany, 1993).
\input{Barin_1993}

\bibitem{Gibbons_PR_1958}
D.~F. Gibbons, \emph{Thermal Expansion of Some Crystals with the Diamond
  Structure}, Phys.\ Rev. \textbf{112}, 136--140 (1958).
\input{Gibbons_PR_1958}

\bibitem{Yim_JAP_1978}
W.~M. Yim and R.~J. Paff, \emph{Thermal expansion of {Al}{N}, sapphire, and
  silicon}, J.\ Appl.\ Phys. \textbf{45}, 1456--1457 (1974).
\input{Yim_JAP_1978}

\bibitem{Okada_JAP_1984}
Y.~Okada and Y.~Tokumaru, \emph{Precise determination of lattice parameter and
  thermal expansion coefficient of silicon between {300} and {1500} {K}}, J.\
  Appl.\ Phys. \textbf{56}, 314--320 (1984).
\input{Okada_JAP_1984}

\bibitem{Chase_AIP_1998}
M.~W. {Chase~Jr.}, \emph{{NIST}-{JANAF} Thermochemical Tables}, Journal of
  Physical and Chemical Reference Data Monographs (American Inst. of Physics,
  1998), fourth edition edn.
\input{Chase_AIP_1998}

\bibitem{Slack_JAP_1975}
G.~A. Slack and S.~F. Bartram, \emph{Thermal expansion of some diamond-like
  crystals}, J.\ Appl.\ Phys. \textbf{46}, 89--98 (1975).
\input{Slack_JAP_1975}

\bibitem{Skinner_AM_1957}
B.~J. Skinner, \emph{The thermal expansion of theoria, periclase, and diamond},
  Am.\ Mineral. \textbf{42}, 39--55 (1957).
\input{Skinner_AM_1957}

\bibitem{Victor_JCP_1962}
A.~C. Victor, \emph{Heat Capacity of Diamond at High Temperatures}, J.\ Chem.\
  Phys. \textbf{36}, 1903--1911 (1962).
\input{Victor_JCP_1962}

\bibitem{McSkimin_JAP_1972}
H.~J. McSkimin and P.~{Andreatch~Jr.}, \emph{Elastic Moduli of Diamond as a
  Function of Pressure and Temperature}, J.\ Appl.\ Phys. \textbf{43},
  2944--2948 (1972).
\input{McSkimin_JAP_1972}

\bibitem{Li_JAP_1986}
Z.~Li and R.~C. Bradt, \emph{Thermal expansion of the hexagonal ({4}{H})
  polytype of {Si}{C}}, J.\ Appl.\ Phys. \textbf{60}, 612--614 (1986).
\input{Li_JAP_1986}

\bibitem{Csonka_PRB_2009}
G.~I. Csonka, J.~P. Perdew, A.~Ruzsinszky, P.~H.~T. Philipsen, S.~Leb\`egue,
  J.~Paier, O.~A. Vydrov, and J.~G. \'Angy\'an, \emph{Assessing the Performance
  of Recent Density Functionals for Bulk Solids}, Phys.\ Rev.\ B \textbf{79},
  155107 (2009).
\input{Csonka_PRB_2009}

\bibitem{Khan_ACTACRIST_1986}
A.~A. Khan, \emph{X-ray determination of thermal expansion of zinc oxide},
  Acta\ Crystallogr.\ Sect.\ A \textbf{24}, 403 (1968).
\input{Khan_ACTACRIST_1986}

\bibitem{Touloukian_NY_1977}
Y.~S. Touloukian, R.~K. Kirby, R.~E. Taylor, and P.~D. Desai,
  \emph{Thermophysical Properties of Matter; Thermal Expansion; Nonmetallic
  Solids}, vol.~13 (IFI-Plenum, New York, 1977).
\input{Touloukian_NY_1977}

\bibitem{Ibach_PSS_1972}
H.~Ibach, \emph{Thermal Expansion of Silicon and Zinc Oxide ({II})}, Phys.\
  Stat.\ Solidi \textbf{33}, 257--265 (1969).
\input{Ibach_PSS_1972}

\bibitem{Kondo_JPNJAP_2008}
S.~Kondo, K.~Tateishi, and N.~Ishizawa, \emph{Structural Evolution of Corundum
  at High Temperatures}, Jpn.\ J.\ Appl.\ Phys. \textbf{47}, 616 (2008).
\input{Kondo_JPNJAP_2008}

\bibitem{Munro_JACERS_1997}
R.~G. Munro, \emph{Evaluated Material Properties for a Sintered
  {$\alpha$}-Alumina}, J.\ Am.\ Ceram.\ Soc. \textbf{80}, 1919--1928 (1997).
\input{Munro_JACERS_1997}

\bibitem{Wachtman_JACERS_1962}
J.~B. {Wachtman~Jr.}, T.~G. Scuderi, and G.~W. Cleek, \emph{Linear Thermal
  Expansion of Aluminum Oxide and Thorium Oxide from {100}$^\circ$ to
  {1100}$^\circ${K}}, J.\ Am.\ Ceram.\ Soc. \textbf{45}, 319--323 (1962).
\input{Wachtman_JACERS_1962}

\bibitem{Madelung_Springer_1999}
O.~Madelung, U.~R\"{o}ssler, and M.~Schulz, eds., \emph{II-VI and I-VII
  Compounds; Semimagnetic Compounds} (Springer Berlin Heidelberg, Berlin,
  Heidelberg, 1999), \doi{10.1007/b71137}.
\input{Madelung_Springer_1999}

\bibitem{Anderson_RGP_1992}
O.~L. Anderson, D.~Isaak, and H.~Oda, \emph{High-temperature elastic constant
  data on minerals relevant to geophysics}, Rev.\ Geophys. \textbf{30}, 57--90
  (1992).
\input{Anderson_RGP_1992}

\bibitem{White_JAP_1996}
G.~K. White and O.~L. Anderson, \emph{Gr\"{u}neisen Parameter of Magnesium
  Oxide}, J.\ Appl.\ Phys. \textbf{37}, 430--432 (1966).
\input{White_JAP_1996}

\bibitem{vanRoekeghem_PRB_2016}
A.~{van Roekeghem}, J.~Carrete, and N.~Mingo, \emph{Anomalous thermal
  conductivity and suppression of negative thermal expansion in {Sc}{F}$_{3}$},
  Phys.\ Rev.\ B \textbf{94}, 020303(R) (2016).
\input{vanRoekeghem_PRB_2016}

\bibitem{kresse_vasp}
G.~Kresse and J.~Hafner, \emph{{\it Ab initio} molecular dynamics for liquid
  metals}, Phys.\ Rev.\ B \textbf{47}, 558--561 (1993).
\input{kresse_vasp}

\bibitem{curtarolo:art104}
C.~E. Calderon, J.~J. Plata, C.~Toher, C.~Oses, O.~Levy, M.~Fornari, A.~Natan,
  M.~J. Mehl, G.~L.~W. Hart, M.~{Buongiorno Nardelli}, and S.~Curtarolo,
  \emph{The {AFLOW} standard for high-throughput materials science
  calculations}, Comput.\ Mater.\ Sci. \textbf{108 Part A}, 233--238 (2015).
\input{curtarolo:art104}

\bibitem{PAW}
P.~E. Bl\"{o}chl, \emph{Projector augmented-wave method}, Phys.\ Rev.\ B
  \textbf{50}, 17953--17979 (1994).
\input{PAW}

\bibitem{PBE}
J.~P. Perdew, K.~Burke, and M.~Ernzerhof, \emph{Generalized Gradient
  Approximation Made Simple}, Phys.\ Rev.\ Lett. \textbf{77}, 3865--3868
  (1996).
\input{PBE}

\bibitem{Perdew_PRL_2008}
J.~P. Perdew, A.~Ruzsinszky, G.~I. Csonka, O.~A. Vydrov, G.~E. Scuseria, L.~A.
  Constantin, X.~Zhou, and K.~Burke, \emph{Restoring the Density-Gradient
  Expansion for Exchange in Solids and Surfaces}, Phys.\ Rev.\ Lett.
  \textbf{100}, 136406 (2008).
\input{Perdew_PRL_2008}

\bibitem{curtarolo:art119}
P.~Nath, J.~J. Plata, D.~Usanmaz, C.~Toher, M.~Fornari, M.~{Buongiorno
  Nardelli}, and S.~Curtarolo, \emph{High throughput combinatorial method for
  fast and robust prediction of lattice thermal conductivity}, Scr.\ Mater.
  \textbf{129}, 88--93 (2017).
\input{curtarolo:art119}

\bibitem{curtarolo:art135}
D.~Hicks, C.~Oses, E.~Gossett, G.~Gomez, R.~H. Taylor, C.~Toher, M.~J. Mehl,
  O.~Levy, and S.~Curtarolo, \emph{{\it AFLOW-SYM}: platform for the complete,
  automatic and self-consistent symmetry analysis of crystals}, Acta\
  Crystallogr.\ Sect.\ A \textbf{74}, 184--203 (2018).
\input{curtarolo:art135}

\bibitem{Wang2010}
Y.~Wang, J.~J. Wang, W.~Y. Wang, Z.~G. Mei, S.~L. Shang, L.~Q. Chen, and Z.~K.
  Liu, \emph{A mixed-space approach to first-principles calculations of phonon
  frequencies for polar materials}, J.\ Phys.:\ Condens.\ Matter \textbf{22},
  202201 (2010).
\input{Wang2010}

\bibitem{Drulis_JAC_2007}
M.~K. Drulis, H.~Drulis, A.~E. Hackemer, A.~Ganguly, T.~{El-Raghy}, and M.~W.
  Barsoum, \emph{On the low temperature heat capacities of {Ti}$_{2}${Al}{N}
  and {Ti}$_{2}${Al}({N}$_{0.5}${C}$_{0.5}$)}, J.\ Alloys\ Compd. \textbf{433},
  59--62 (2007).
\input{Drulis_JAC_2007}

\bibitem{Jung_JAC_2010}
D.~Y. Jung, K.~Kurosaki, C.~Kim, H.~Muta, and S.~Yamanaka, \emph{Thermal
  expansion and melting temperature of the half-Heusler compounds:
  {$M$}{Ni}{Sn} ({$M$}={Ti}, {Zr}, {Hf})}, J.\ Alloys\ Compd. \textbf{489},
  328--331 (2010).
\input{Jung_JAC_2010}

\bibitem{Lane_JCCERS_2011}
N.~J. Lane, S.~C. Vogel, and M.~W. Barsoum, \emph{Temperature-Dependent Crystal
  Structures of {Ti}$_{2}${Al}{N} and {Cr}$_{2}${Ge}{C} as Determined from High
  Temperature Neutron Diffraction}, J.\ Am.\ Ceram.\ Soc. \textbf{94},
  3473--3479 (2011).
\input{Lane_JCCERS_2011}

\bibitem{Aliev_PBCM_1990}
F.~G. Aliev, V.~V. Kozyrkov, V.~V. Moshchalkov, R.~V. Scolozdra, and
  K.~Durczewski, \emph{Narrow band in the intermetallic compounds {$M$NiSn
  ({$M$}={Ti}, {Zr}, {Hf})}}, Z.\ Phys.\ B:\ Condens.\ Matter \textbf{80},
  353--357 (1990).
\input{Aliev_PBCM_1990}

\bibitem{curtarolo:art121}
M.~J. Mehl, D.~Hicks, C.~Toher, O.~Levy, R.~M. Hanson, G.~L.~W. Hart, and
  S.~Curtarolo, \emph{The {AFLOW} Library of Crystallographic Prototypes: Part
  1}, Comput.\ Mater.\ Sci. \textbf{136}, S1--S828 (2017).
\input{curtarolo:art121}

\bibitem{Hicks_arXiv_2018}
D.~Hicks, M.~J. Mehl, E.~Gossett, C.~Toher, O.~Levy, R.~M. Hanson, G.~Hart, and
  S.~Curtarolo, \emph{The {AFLOW} Library of Crystallographic Prototypes: {Part
  2}}, arXiv:1806.07864 [cond-mat.mtrl-sci]  (2018).
\input{Hicks_arXiv_2018}

\bibitem{A_cF8_227_a}
\url{http://aflow.org/CrystalDatabase/A_cF8_227_a.html}.
\input{A_cF8_227_a}

\bibitem{AB_hP4_186_b_b}
\url{http://aflow.org/CrystalDatabase/AB_hP4_186_b_b.html}.
\input{AB_hP4_186_b_b}

\bibitem{A2B3_hR10_167_c_e}
\url{http://aflow.org/CrystalDatabase/A2B3_hR10_167_c_e.html}.
\input{A2B3_hR10_167_c_e}

\bibitem{AB_cF8_225_a_b}
\url{http://aflow.org/CrystalDatabase/AB_cF8_225_a_b.html}.
\input{AB_cF8_225_a_b}

\bibitem{AB_cP2_221_b_a}
\url{http://aflow.org/CrystalDatabase/AB_cP2_221_b_a.html}.
\input{AB_cP2_221_b_a}

\bibitem{ABC_cF12_216_b_c_a}
\url{http://aflow.org/CrystalDatabase/ABC_cF12_216_b_c_a.html}.
\input{ABC_cF12_216_b_c_a}

\bibitem{ABC2_hP8_194_d_a_f}
\url{http://aflow.org/CrystalDatabase/ABC2_hP8_194_d_a_f.html}.
\input{ABC2_hP8_194_d_a_f}

\bibitem{Togo_prb_2015}
A.~Togo, L.~Chaput, and I.~Tanaka, \emph{Distributions of phonon lifetimes in
  {Br}illouin zones}, Phys.\ Rev.\ B \textbf{91}, 094306 (2015).
\input{Togo_prb_2015}

\end{thebibliography}
\end{document}